%% file: main.tex
\documentclass[sigconf,bookmarks=false,screen=true]{acmart}
\pdfoutput=1

\input{header.tex}

\title{A Longitudinal Analysis of Bloated Java Dependencies}

\author{C\'esar Soto-Valero}
\affiliation{%
    \institution{KTH Royal Institute of Technology, Sweden}
}
\email{cesarsv@kth.se}

\author{Thomas Durieux}
\affiliation{%
    \institution{KTH Royal Institute of Technology, Sweden}
}
\email{thomas@durieux.me}

\author{Benoit Baudry}
\affiliation{%
    \institution{KTH Royal Institute of Technology, Sweden}
}
\email{baudry@kth.se}

\begin{document}

\begin{abstract}

We  study the evolution and impact of bloated dependencies in a single software ecosystem: Java/Maven. Bloated dependencies are third-party libraries that are packaged in the application binary but are not needed to run the application. 
We analyze the history of \nbProjects Java projects. This historical data includes \nbDependencies distinct dependencies, which we study across a total of \nbCommits versions of Maven dependency trees.
Bloated dependencies steadily increase over time, and \ProbBloatedDirect of the  direct dependencies that are bloated  remain bloated in all subsequent versions of the studied projects. This empirical evidence suggests that developers can safely remove a bloated dependency. We further report novel insights regarding the unnecessary maintenance efforts induced by bloat. We find that \np[\%]{22} of dependency updates performed by developers are made on bloated dependencies, and that Dependabot suggests a similar ratio of updates on bloated dependencies.
\end{abstract}


\maketitle

\input{body.tex}

\bibliography{references}

\end{document}

%% file: header.tex
\copyrightyear{2021} 
\acmYear{2021} 
\setcopyright{rightsretained} 
\acmConference[ESEC/FSE'21, August 23--27, 2021 ]{ESEC/FSE'21}{Athens, Greece}{Virtual Event}
\acmBooktitle{ESEC/FSE'21}
\acmDOI{nnnnnnn.nnnnnnn}
\acmISBN{nnnnnnn.nnnnnnn}

\usepackage[T1]{fontenc}
\usepackage{url}
\usepackage[keeplastbox]{flushend}
\usepackage{xspace}
\usepackage{caption}
\usepackage[inline]{enumitem}
\usepackage{ifthen}
\usepackage{balance}
\usepackage{microtype}
\usepackage[autolanguage]{numprint}
\newcommand*\np[2][z]{
\ifx z#1%
$\numprint{#2}$%
\else%
$\numprint[#1]{#2}$%
\fi\xspace%
}

\usepackage{graphicx}
\usepackage{subcaption}
\usepackage{pifont}

\usepackage{booktabs}
\usepackage{multirow}
\usepackage{threeparttable}
\usepackage{xcolor}
\usepackage{array,makecell}
\usepackage{diagbox}
\usepackage{tabularx}

\usepackage{mathtools}
\usepackage{amsmath}

\usepackage{tcolorbox}
\usepackage{pgfplots}
\usepackage{pgfplotstable}

\usepackage{enumitem}
\usepackage[framemethod=TikZ]{mdframed}
\usepackage[scaled]{beramono} 

\usepgfplotslibrary{statistics}
\usetikzlibrary{calc,patterns,fpu} 
\pgfplotsset{compat=1.14} 
\usepackage{silence}
\pgfplotsset{
    /pgf/declare function={
        Floor(\x) = round(\x-0.49);
    },
    show sum on top/.style={
        /pgfplots/scatter/@post marker code/.append code={%
            \path let \p1=($(normalized axis cs:%
                        \pgfkeysvalueof{/data point/x},%
                        \pgfkeysvalueof{/data point/y})%
                        -(normalized axis cs:\pgfkeysvalueof{/data point/x},0)$)
            in node[
                at={(normalized axis cs:%
                        \pgfkeysvalueof{/data point/x},%
                        \pgfkeysvalueof{/data point/y})%
                },
                anchor={-90*sign(\y1)},yshift={sign(\y1)*2pt}
            ]
            {\pgfmathprintnumber{\pgfkeysvalueof{/data point/y}}};
        },
    }
}

\definecolor{blau_1a}{RGB}{93,133,195}
\definecolor{blau_2a}{RGB}{0,156,218}
\definecolor{gruen_3a}{RGB}{80,182,149}
\definecolor{gruen_4a}{RGB}{175,204,80}
\definecolor{gruen_5a}{RGB}{221,223,72}
\definecolor{orange_6a}{RGB}{255,224,92}
\definecolor{orange_7a}{RGB}{248,186,60}
\definecolor{rot_8a}{RGB}{238,122,52}
\definecolor{rot_9a}{RGB}{233,80,62}
\definecolor{lila_10a}{RGB}{201,48,142}
\definecolor{lila_11a}{RGB}{128,69,151}

\definecolor{blau_1b}{RGB}{0,90,169}
\definecolor{blau_2b}{RGB}{0,131,204}
\definecolor{gruen_3b}{RGB}{0,157,129}
\definecolor{gruen_4b}{RGB}{153,192,0}
\definecolor{gruen_5b}{RGB}{201,212,0}
\definecolor{orange_6b}{RGB}{253,202,0}
\definecolor{orange_7b}{RGB}{245,163,0}
\definecolor{rot_8b}{RGB}{236,101,0}
\definecolor{rot_9b}{RGB}{230,0,26}
\definecolor{lila_10b}{RGB}{166,0,132}
\definecolor{lila_11b}{RGB}{114,16,133}

\newboolean{showcomments}
\setboolean{showcomments}{true}

\input{ChartInTable.tex}

\input{macros.tex}

\definecolor{mygray}{RGB}{240,240,240}

\newcommand{\depclean}{\textsc{DepClean}\xspace}
\newcommand{\pom}{\texttt{pom.xml}\xspace}

\ifthenelse{\boolean{showcomments}}
 { \newcommand{\mynote}[2]{
      \fbox{\bfseries\sffamily\scriptsize#1}
        {\small$\blacktriangleright$\textsf{\textcolor{red}{{\em #2}\bf }}$\blacktriangleleft$}}}
        { \newcommand{\mynote}[2]{}}

\definecolor{eminence}{RGB}{108,48,130}
\definecolor{weborange}{RGB}{255,165,0}
\definecolor{frenchplum}{RGB}{129,20,82}
\definecolor{darkgreen}{RGB}{10, 92, 10}

\mdfdefinestyle{mpdframe}{
    nobreak                     =true,
    backgroundcolor             =black!3,
    frametitlebackgroundcolor   =black!15,
    frametitlerule              =false,
    roundcorner                 =5pt,
    innermargin                 =0.2cm,
    innerleftmargin             =0.2cm,
    innerrightmargin            =0.2cm,
    innertopmargin              =0.2cm,
    innerbottommargin           =0.2cm,
    shadowsize                  =1pt
}

%% file: ChartInTable.tex
\usepackage{fp}
\newcommand{\ShowAbsoluteNumber}[1]{%
\ifnum #1<10%
{\hspace*{0pt}#1}%
\else%
\ifnum #1<100%
{\hspace*{0pt}#1}%
\else%
\ifnum #1<1000%
{\hspace*{0pt}#1}%
\else%
{\numprint{#1}}%
\fi%
\fi%
\fi%
}

\newcommand{\ShowPercentage}[2]{%
\FPeval\percentage{round(#1/#2*100,0)}%
\FPeval\percentageOneDecimal{round(#1/#2*100,1)}%
\ifnum \percentage=0%
{\np[\%]{\FPprint{percentageOneDecimal}}}%
\else%
\ifnum \percentage<10%
{\np[\%]{\FPprint{percentageOneDecimal}}}%
\else%
{\np[\%]{\FPprint{percentageOneDecimal}}}%
\fi%
\fi%
\xspace
}
\newlength\BARSIZE  
\newcommand{\inlinechart}[2]{%
\FPeval{\BLACKBARSIZE}{#1/#2}\textcolor{black!80}{\rule{\BLACKBARSIZE\BARSIZE}{1.6ex}}%
\FPeval{\BLACKBARSIZE}{1 - (#1/#2)}\textcolor{black!10}{\rule{\BLACKBARSIZE\BARSIZE}{1.6ex}}%
}

\newcommand*\percent[3][v]{%
\ifx q#1%
    \np{#2}/\np{#3}(\ShowPercentage{#2}{#3})\else%
\ifx p#1%
    \np{#2}(\ShowPercentage{#2}{#3})\else%
\ifx c#1%
    \inlinechart{#2}{#3}%
\else%
    \np{#2}%
    \ifx r#1%
        /\np{#3}%
    \fi%
    \hspace*{0.5ex}(\ShowPercentage{#2}{#3}) %
    \inlinechart{#2}{#3}%
    \xspace
\fi\fi\fi%
}

%% file: macros.tex
\def\nbTotalProject{147991}
\def\nbTotalProjectStr{\np{\nbTotalProject}}

\def\nbDatasetProjectsNum{500}
\def\nbDatasetReleasesNum{47276}
\def\nbDatasetDependabotNum{2017}
\FPeval{nbDatasetCommitsNum}{round(\nbDatasetReleasesNum + \nbDatasetDependabotNum, 0)}

\def\nbDatasetProjects{\np{\nbDatasetProjectsNum}}
\def\nbDatasetReleases{\np{\nbDatasetReleasesNum}}
\def\nbDatasetDependabot{\np{\nbDatasetDependabotNum}}
\def\nbDatasetCommits{\np{\nbDatasetCommitsNum}}

\def\repo{\url{https://github.com/castor-software/longitudinal-bloat}}

\def\nbProjectsNum{435}
\def\nbReleasesNum{29822}
\def\nbDependabotNum{1693}
\def\nbDependenciesNum{48469}
\def\nbBloatedDependenciesNum{40493}
\def\nbUsedUpdateNum{25124}
\def\nbBloatUpdateNum{5375}
\def\nbDepUsedUpdateNum{9403}
\def\nbDepBloatUpdateNum{2659}
\def\nbUsedDependabotUpdateNum{2452}
\def\nbBloatDependabotUpdateNum{716}
\def\nbBloatIntroducedByUpdateNum{1883}
\def\originTransitiveBloatNewDepNum{33370}
\def\originTransitiveBloatRemoveNum{206}
\def\originTransitiveBloatUpdateNum{495}
\def\originTransitiveBloatNewVersionNum{124}
\def\originDirectBloatNewDepNum{1868}
\def\originDirectBloatRemoveNum{194}
\def\originDirectBloatUpdateNum{153}
\def\originDirectBloatNewVersionNum{47}
\def\NbOriginBloatDependencyNum{25359}
\def\NbOriginBloatDirectDependencyNum{1987}
\def\NbOriginBloatTransitiveDependencyNum{23442}

\FPeval{nbUpdatesNum}{round(\nbUsedUpdateNum + \nbBloatUpdateNum + \nbUsedDependabotUpdateNum + \nbBloatDependabotUpdateNum, 0)}
\FPeval{nbDepUpdatesNum}{round(\nbDepUsedUpdateNum + \nbDepBloatUpdateNum + \nbUsedDependabotUpdateNum + \nbBloatDependabotUpdateNum, 0)}
\FPeval{nbDepDevUpdatesNum}{round(\nbDepUsedUpdateNum + \nbDepBloatUpdateNum, 0)}
\FPeval{nbHumanUpdatesNum}{round(\nbUsedUpdateNum + \nbBloatUpdateNum, 0)}
\FPeval{nbDependabotUpdatesNum}{round(\nbUsedDependabotUpdateNum + \nbBloatDependabotUpdateNum, 0)}
\FPeval{nbTotalBloatUpdatesNum}{round(\nbBloatUpdateNum + \nbBloatDependabotUpdateNum, 0)}
\FPeval{nbCommitsNum}{round(\nbReleasesNum + \nbDependabotNum, 0)}

\FPeval{nbTotalProjectsWithDependabotUpdates}{round(\nbDepDevUpdatesNum + \nbDependabotUpdatesNum, 0)}

\FPeval{totalOriginDirect}{round(\originDirectBloatNewDepNum + \originDirectBloatRemoveNum + \originDirectBloatUpdateNum, 0)}
\FPeval{totalOriginTransitive}{round(\originTransitiveBloatNewDepNum + \originTransitiveBloatRemoveNum + \originTransitiveBloatUpdateNum, 0)}

\def\nbProjects{\np{\nbProjectsNum}}

\def\nbCommits{\np{\nbCommitsNum}}
\def\nbDependencies{\np{\nbDependenciesNum}}

\def\ProbBloatedDirect{\np[\%]{89.2}}
\def\ProbBloatedTransitive{\np[\%]{93.3}}

\newcommand{\ie}{i.e.\@\xspace}

\newcommand{\eg}{e.g.\@\xspace}
\newcommand{\etal}{et al.\@\xspace}

\newcommand{\wrt}{w.r.t.\@\xspace}

%% file: body.tex
\section{Introduction}

Software is bloated. From single Unix commands \cite{Holzmann2015} to web browsers \cite{Qian2020}, most applications embed a part of code that is unnecessary to their correct operation. Several debloating tools have emerged in recent years~\cite{Sharif2018,Qian2020,Valero2020,Jiang2016,Rastogi2017,Qian2019} to address the security and maintenance issues posed by excessive code at various granularity levels. However, these works do not analyze the evolution of bloat over time. Understanding software bloat in the perspective of software evolution \cite{Hilton0M18,Spinellis17,TeytonFPB14} is crucial to promote debloating tools towards software developers. In particular, developers, when proposed to adapt a debloating tool, wonder if a piece of bloated code might be needed in coming releases, or what is the actual issue with bloat.\looseness=-1

This work proposes the first longitudinal analysis of software bloat. We focus on bloat among software dependencies \cite{Cox2019,Gustavsson20,BoldiG21, Valero2019} in the Java/Maven ecosystem. 
Bloated dependencies are software libraries that are unnecessarily part of software projects, \ie, when the dependency is removed from the project, it still builds successfully.
In previous work \cite{Valero2020}, we showed that the Maven ecosystem is permeated with bloated dependencies, and that they are present even in well maintained Java projects. Our study revealed that software developers are keen on removing bloated dependencies, but that removing code is a complex socio-technical decision, which benefits from solid evidence about the actual benefits of debloating.

Motivated by these observations about bloated dependencies, we conduct a large scale empirical study about the evolution of these dependencies in Java projects. We analyze the emergence of bloat, the evolution of the dependencies statuses, and the impact of bloat on maintenance.
We have collected a unique dataset of \nbCommits versions of dependency trees  from \nbProjects open-source Java projects.
Each version of a tree is a snapshot of one project's dependencies, for which we determine a status, \ie bloated or used.  
We rely on \depclean, the state-of-the-art tool to detect bloated dependencies in Maven projects. 
We analyze the evolution of  \nbDependencies distinct dependencies per project and we observe that  \percent[q]{\nbBloatedDependenciesNum}{\nbDependenciesNum} of them are bloated at one point in time, in our dataset.

Our longitudinal analysis of bloated Java dependencies investigates both the evolution of bloat, as well as the relation between bloat and regular maintenance activities such as dependency updates.
We present original quantitative results regarding the evolution of bloated dependencies. We first show a clear increasing trend in the number of bloated dependencies.
Next, we investigate how the usage status of dependencies evolves over time.
This analysis is a key contribution of our work where we demonstrate that a dependency that is bloated is very likely to remain bloated over subsequent versions of a project.
We present the first observations about the impact of regular maintenance activities on software bloat.
Besides, we analyze the impact of Dependabot, a popular dependency management bot, on these activities. 
We show that developers regularly update bloated dependencies, and that many of these updates are suggested by Dependabot.
Furthermore, we systematically investigate the root of the bloat emergence,  and find that \ShowPercentage{\originDirectBloatNewDepNum}{\totalOriginDirect} of the bloated dependencies are bloated as soon as they are added in the dependency tree of a project.

\begin{figure*}[!t]
\minipage[t]{0.33\textwidth}%
    \includegraphics[origin=c,width=0.9\textwidth]{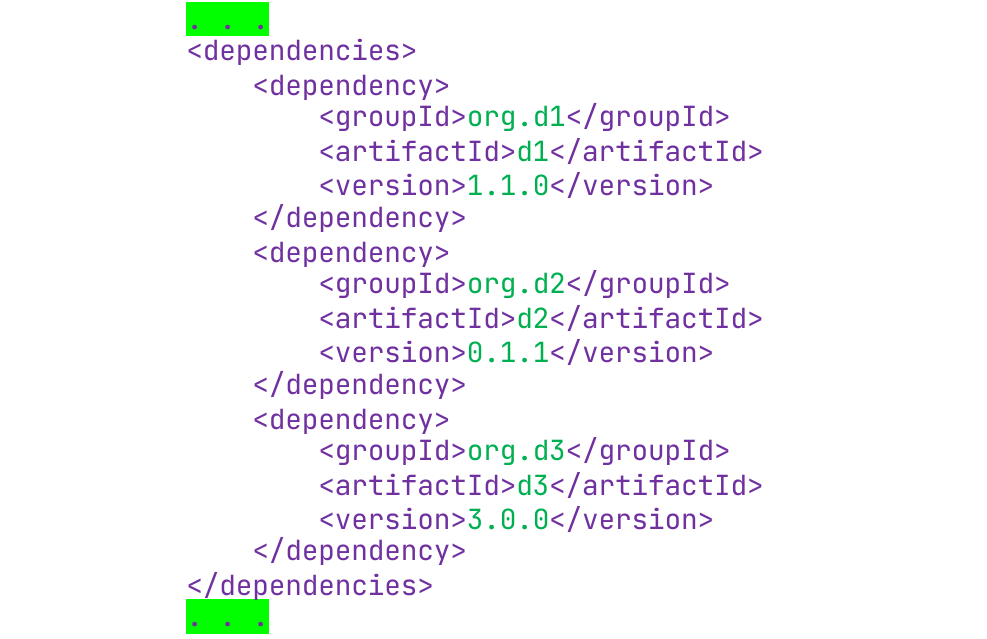}
    \caption{Dependency declaration.}
    \label{fig:pom}
\endminipage\hfill
\minipage[t]{0.33\textwidth}%
    \includegraphics[origin=c,width=0.9\textwidth]{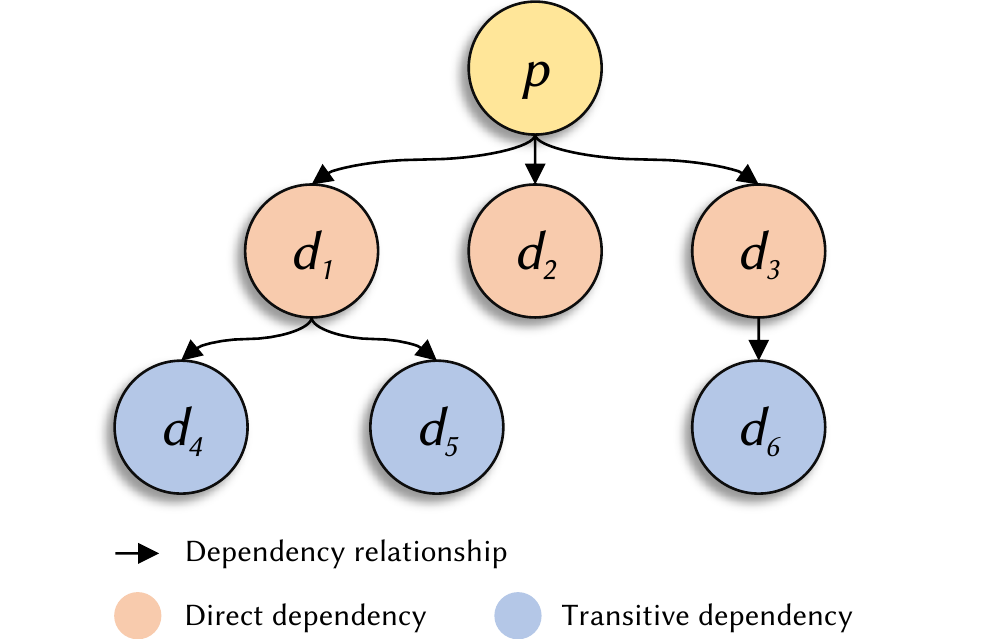}
    \caption{Dependency tree.}
    \label{fig:dt}
\endminipage\hfill
\minipage[t]{0.33\textwidth}%
    \includegraphics[origin=c,width=0.9\textwidth]{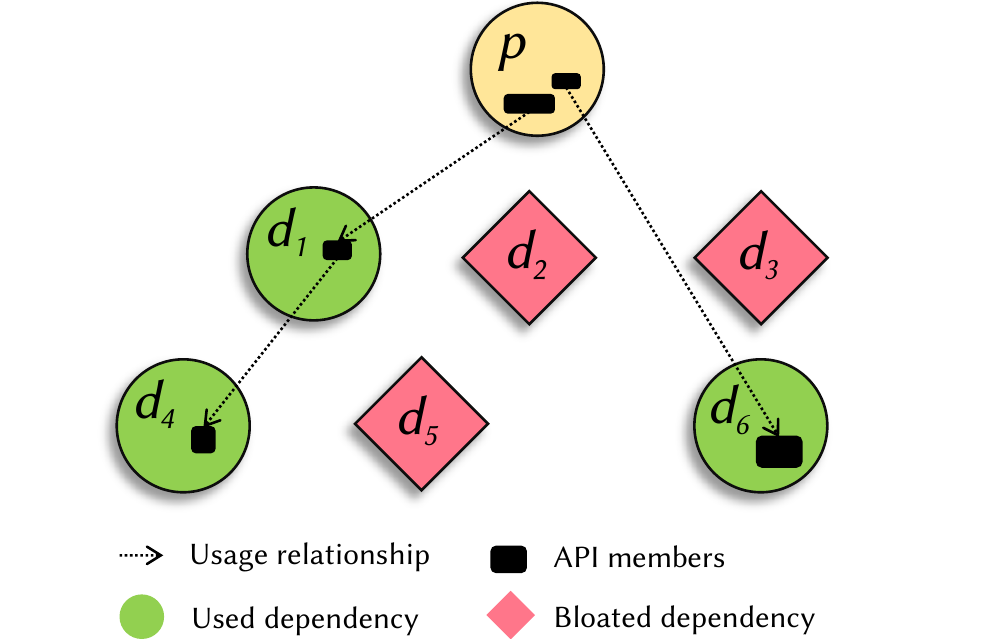}
    \caption{Dependency usage.}
    \label{fig:du}
\endminipage
\end{figure*}

To summarize, the contributions of this paper are:
\begin{itemize}
    \item A longitudinal analysis of software dependencies' usage in \nbCommits versions of Maven dependency trees. Our results confirm the generalized presence of bloated dependencies and show their increase over time. 
    \item A quantitative analysis of the stability of bloated dependencies: \ProbBloatedDirect of direct dependencies remain bloated. This is a concrete insight that motivates debloating dependencies. 
    \item A novel analysis of unnecessary updates on bloated dependencies made by a software bot. We find that developers often accept Dependabot's suggestions without considering if the dependency is actually used or not.
    \item A qualitative manual analysis of the origin of bloated dependencies, that reveals that adding dependencies is the principal reason that originates this type of software bloat. 
\end{itemize}

\section{Background}

\begin{figure*}[t]
  \centering
  \includegraphics[origin=c,width=0.9\textwidth]{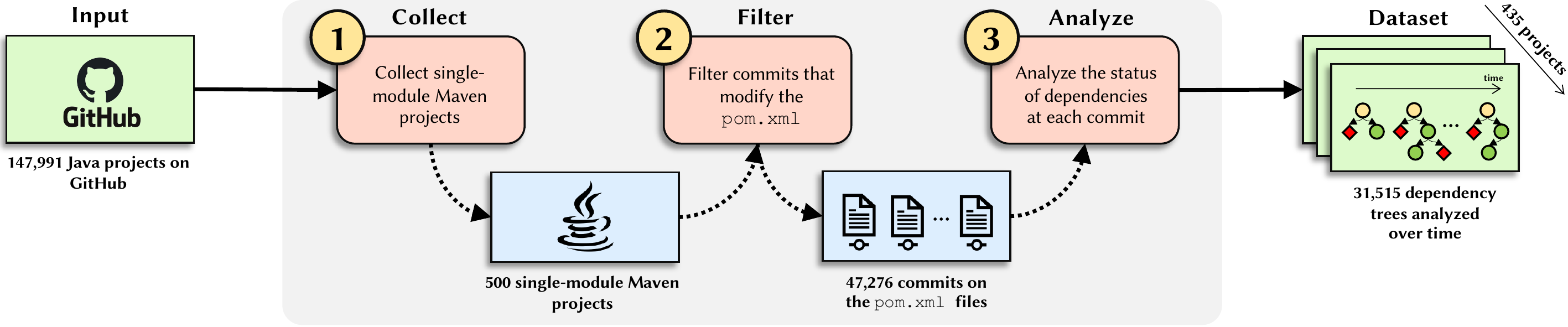}
  \caption{Overview of our data collection pipeline. From a set of \nbTotalProjectStr Java projects on GitHub, we analyze the usage status of the dependencies in \nbProjects Maven projects over time, to produce a dataset of \nbCommits dependency trees.}
  \label{fig:overview}
\end{figure*}

In this work, we consider a software project as a collection of Java source code files and configuration files organized to be build with Maven.\footnote{\url{https://maven.apache.org}}
In this section, we present the key concepts for the analysis of a project $p$ in the context of the set of its software dependencies, denoted as $\mathcal{D}$.

\begin{definition}
\textbf{Maven dependency: }
A Maven dependency defines a relationship between a project $p$ and another compiled project $d \in \mathcal{D}$. 
Dependencies are  compiled JAR files, uniquely identified with a triplet (\texttt{G:A:V}) where \texttt{G} is the \texttt{groupId}, \texttt{A} is the \texttt{artifactId}, and \texttt{V} is the \texttt{version}.
Dependencies are defined within a scope, which determines at which phase of the Maven build cycle the dependency is required (\eg, \texttt{compile}, \texttt{test}, \texttt{runtime}).
\end{definition}

A Maven project declares a set of direct  dependencies in a specific configuration file known as \pom (acronym for “Project Object Model”), located at the root of the project.
\autoref{fig:pom} shows an excerpt of the dependency declaration in the \pom of a project $p$.
In this example, developers explicitly declare the usage of three dependencies: $d_1$, $d_2$, and $d_3$.
Note that the \pom of a Maven project is a configuration file subject to constant change and evolution: developers usually commit changes to add, remove, or update the version of a dependency.

\begin{definition}
\textbf{Direct dependency: }
The set of direct dependencies $\mathcal{D}_\text{direct} \subset \mathcal{D}$ of a project $p$ is the set of dependencies declared in $p$'s \pom file. 
Direct dependencies are  declared in the \pom by the developers, who explicitly manifest the intention of using the dependency.
\end{definition}

\begin{definition}
\textbf{Transitive dependency:}
The set of transitive dependencies $\mathcal{D}_\text{transitive} \subset \mathcal{D}$ of a project $p$ 
is the set of dependencies obtained from the transitive closure of direct dependencies. 
Transitive dependencies are resolved automatically by Maven, which means that developers do not need to explicitly declare these dependencies.
\end{definition}

\begin{definition}
\textbf{Dependency tree: }
The dependency tree of a Maven project $p$ is a direct acyclic graph of the dependencies of $p$, where $p$ is the root node and the edges represent dependency relationships between $p$ and the dependencies in $\mathcal{D}$.
\end{definition}

To construct the dependency tree, Maven relies on its specific dependency resolution mechanism.\footnote{\url{https://maven.apache.org/guides/introduction/introduction-to-dependency-mechanism.html}}
First, Maven determines the set of declared dependencies based on the \pom file of the project. 
Then, it fetches the JARs of the dependencies that are not present locally from external repositories, \eg, Maven Central.\footnote{\url{https://mvnrepository.com/repos/central}} 

\autoref{fig:dt} illustrates the dependency tree of the project $p$, which \pom file is in \autoref{fig:pom}.
The project has three direct dependencies, as declared in its \pom, and three transitive dependencies, as a result of the Maven dependency resolution mechanism.
$d_4$ and $d_5$ are induced transitively from $d_1$, whereas the transitive dependency $d_6$ is induced from $d_3$.
Note that all the bytecode of these transitive dependencies is present in the classpath of project $p$, and hence they will be packaged with it, whether or not they are actually used by $p$.

\begin{definition}
\textbf{Bloated dependency: }
A dependency $d \in \mathcal{D}$ in a software project $p$ is said to be bloated if there is no path in the dependency tree of $p$, between $p$ and $d$, such that none of the elements in the API of $d$ are used, directly or indirectly, by $p$. 
\end{definition}

We introduced the concept of bloated dependencies in 2020~\cite{Valero2020}.
Although they are present in the dependency tree of software projects, bloated dependencies are useless and, therefore, developers can consider removing them.

\begin{definition}
\textbf{Dependency usage status: }
The usage status of a dependency $d \in \mathcal{D}$ determines if $d$ is \textit{used} or \textit{bloated} \wrt to $p$, at a specific time of the development of $p$.
\end{definition}

\autoref{fig:du} shows an hypothetical example of the dependency usage tree of project $p$.
Suppose that $p$ directly calls two sets of instructions in the direct dependency $d_1$ and the transitive dependency $d_6$.
Then, the subset of instructions called in $d_1$ also calls instructions in $d_4$.
In this case, the dependencies $d_1$, $d_4$, and $d_6$ are used by $p$, while dependencies $d2$, $d3$, and $d5$ are bloated dependencies.

Figures \ref{fig:pom}, \ref{fig:dt}  and \ref{fig:du} illustrate the status of a project's dependencies at some point in time. Yet, the \pom file, the dependency tree, and the status of dependencies are prone to change for several reasons. 
For example, a dependency that was used can become bloated after a dependency migration or after some refactoring activities that remove the usage link between the project and some of its dependencies. It is also possible that developers add dependencies in the \pom file or that more transitive dependencies appear in the tree, e.g., when updating the direct dependencies.
This work investigates these software evolution changes and their impact on bloat and maintenance.

\section{Study Design} 

In this section, we present the research protocols that we follow to conduct our empirical study, including the research questions (RQs), the tooling utilized to detect bloated dependencies, the data collection, and our methodology to address each RQ. 

\subsection{Research Questions}\label{sec:rqs}

In this paper, we study four different aspects of bloated dependencies. Our analysis is guided by the following research questions.
\vspace{-1em}
\begin{itemize}[leftmargin=24pt]
    \item[\textbf{RQ1}.] \textbf{How does the amount of bloated dependencies evolve across releases?}
    With this first question, we aim at consolidating the body of knowledge about software bloat. 
    Several recent studies have shed light on the massive presence of bloat in different types of software projects~\cite{Jiang2016,Bruce2020,Qian2019,Sharif2018,Rastogi2017}. The growth of bloat is an important motivation for these works. Yet, this growth has never been assessed nor quantified. Our first research question addresses this lack, analyzing the evolution of the amount of bloat over time. 
    \item[\textbf{RQ2}.] \textbf{How does the usage status of each dependency evolve across time?}
    Tools that remove bloated code are designed under the assumption that a piece of code that is bloated at some point in time will always be bloated, hence it makes sense to remove it.
    In this second research question, we investigate whether this assumption holds true in the case of bloated Java dependencies. We analyze how the usage status of dependencies evolves over time, from used to bloated, or vice versa. 
    \item[\textbf{RQ3}.] \textbf{Do developers maintain dependencies that are bloated?}
    Bloated dependencies needlessly waste time and resources, \eg, space on disk, build time, performance. 
    However, one of the major issues related to this type of dependency is the unnecessary maintenance effort.
    In this research question, we investigate how often developers modify the \pom to update  dependencies that are actually bloated.
    \item[\textbf{RQ4}.] \textbf{What development practices change the usage status of dependencies?}
    The emergence of bloat is due to various code maintenance activities, \eg, refactoring the code, or modifying the \pom.
    In this research question, we expand the quantitative analysis of the status of each dependency and perform an in-depth analysis of the causes of dependency bloat.
\end{itemize}
\vspace{-1em}

\subsection{Detection of Bloated Dependencies}

To analyze the status of dependencies of Maven projects, we rely on \depclean.\footnote{\url{https://github.com/castor-software/depclean}}
This is an open-source tool that implements a practical way of detecting bloated dependencies in the complete dependency tree of a Java Maven project.
\depclean runs a static analysis, at the bytecode level, to detect the usage of direct and transitive dependencies.
To do so, \depclean constructs a static call-graph of API members' calls among the bytecode of the project and its dependencies.
Then, it determines which dependencies are referenced, either directly by the project or indirectly via transitive dependencies.
If none of the API members of a dependency are referenced, \depclean reports the dependency as bloated, \ie, the dependency is not necessary to build the project.
\depclean generates a report with the status of each dependency, a list of API members that are used at least once, for each used dependency. The tool also generates a modified version of the \pom without bloated dependencies. 

\subsection{Data Collection}\label{sec:data_collection}

The dataset used in our study consists of a collection of subsequent versions of Maven dependency trees \cite{durieux21}.
Each dependency in these trees is analyzed in order to determine its status: used or bloated.
\autoref{fig:overview} summarizes the process we follow to collect this dataset.
Rounded rectangles represent procedures, non-rounded rectangles represent intermediate data results.

\ding{182} \textbf{\textit{Collect.}}
Our data collection pipeline starts from the list of Java projects extracted from GitHub by Loriot~\etal~\cite{styler}. 
The authors queried the GitHub API on June 9th of 2020, and provide a list of GitHub URLs including all projects that use Java as the primary programming language. 
From this list, we keep only projects with more than $5$ stars.
This initial dataset contains a total of \nbTotalProjectStr Java projects.
Then, we inspect the projects' files and select those containing a single \pom file in the root of the repository, to focus our longitudinal analysis on  single-module Maven projects. 
This first data collection step provides a set of \np{34560} Java projects.

\ding{183} \textbf{\textit{Filter.}}
In this second step, we check all the commits on the \pom file  to determine the version of the project declared in the \pom.
Each time the version of the project changes and it is not a SNAPSHOT or a beta-version, we consider that the commit represents a new release.
We sort the list of projects by the number of releases and then we select the first \nbDatasetProjects projects.
We focus on release commits since a release represents a stable version of the project, which is a suitable moment to consider the presence of bloated dependencies.
In addition to the project releases, we collect the commits that have been created by Dependabot,\footnote{\url{https://dependabot.com}} a popular software bot that automatizes the update of dependencies on GitHub~\cite{Dey2020}.
The goal is to  determine how many bloated dependencies have been updated as a result of a pull request not made by a human.
We identify \nbDatasetDependabot Dependabot commits for \percent[q]{143}{\nbDatasetProjectsNum} projects.
At the end of this step, we have a total of \nbDatasetProjects projects, as well as \nbDatasetCommits commits, including \nbDatasetReleases release commits.

\ding{184} \textbf{\textit{Analyze.}}
The final and most complex step in our pipeline is to analyze the status of dependencies in the \nbDatasetCommits commits.
We perform the following tasks: 
1) clone the repository and checkout the commit,
2) compile the project using Maven,
3) if the project compiles, then we execute \depclean on the commit to obtain the dependency usage status. We analyze dependencies that have a \texttt{compile} or \texttt{test} scope. 
The compilation task is the most crucial and difficult task because it involves downloading dependencies, having the correct version of Java and having a proper project state, \ie, the Java code needs to be valid.
We mitigate those problems by compiling the projects with Java 11 and then with Java 8.
By trying to compile with Java 8 when the project does not compile with Java 11, we increase the number of successful compilations by around \np[\%]{20}.
We also use a proxy for Maven that caches and looks for dependencies in five different repositories to increase the chances to resolve them.
In total, the proxy cached \np{198611} dependencies and \np[Gb]{165} of data.
As side effects, the proxy speeds up the resolution of dependencies and increases the reproducibility of the study, \ie, Maven will always resolve the same dependencies even if we recompile the projects after several years.

This final step of our pipeline outputs the definitive dataset for our longitudinal study:  the dependency usage trees of \percent[p]{\nbCommitsNum}{\nbDatasetCommitsNum} commits collected from \percent[p]{\nbProjectsNum}{\nbDatasetProjectsNum} projects. These trees capture the history of \nbDependencies dependency relationships, including \np{\NbOriginBloatDirectDependencyNum} direct dependencies and  \np{\NbOriginBloatTransitiveDependencyNum} transitive dependencies. 
Among the commits,  \percent[p]{\nbReleasesNum}{\nbDatasetReleasesNum}  are project releases and \percent[p]{\nbDependabotNum}{\nbDatasetDependabotNum} are Dependabot commits. 
We have kept only the projects for which we can successfully analyze at least two dependency tree versions.
The dataset consists of a \texttt{JSON} file per commit for each project, containing the status of each dependency at every point in time. 
The dataset and the scripts are available in our experiment repository.\footnote{\repo}

\input{tables/ts_descriptive}

\autoref{ts_descriptive} shows descriptive statistics of our dataset. 
The \nbProjectsNum~projects have been active for periods ranging from five months to \np{235} months (12 years and 7 months), with most of them in the range \np{48.5} months (1st Qu.) to \np{109.5} months (3rd Qu.).
The number of dependency trees analyzed for each project ranges from \np{2} to \np{819} (Median = \np{58}, 1st Qu. = \np{41}, 3rd Qu. = \np{79}).
The table also reports  the number of direct dependencies in the oldest analyzed commit (Median = \np{5}, 1st Qu. = \np{3}, 3rd Qu. = \np{10}), and transitive dependencies (Median = \np{10.5}, 1st Qu. = \np{2}, 3rd Qu. = \np{56}).
The last two lines in the table give the number of direct dependencies in the most recent analyzed commit (Median = \np{10}, 1st Qu. = \np{5}, 3rd Qu. = \np{18}), and transitive dependencies (Median = \np{25}, 1st Qu. = \np{6.5}, 3rd Qu. = \np{82.5}).

\subsection{Methodology for RQ1}\label{sec:methodology_rq1}



In RQ1, we analyze the evolution of the number of bloated dependencies over time.
We start with a global analysis of the bloat trend in direct and transitive dependencies.
To do so, we aggregate the total number of bloated dependencies in all projects on a monthly basis and compute the average values.
Next, we look at each project separately and assign a bloat evolution trend to each of them.
We represent the number of dependencies at each commit in a project as a time series.
Let $p$ be a Maven project, $\mathcal{B}_p = b_1, b_2, ...b_n$ represents a time series of length $n$. 
A time step in this series represents one commit that modifies the \pom of $p$.
Each $b_i$ is the total number of bloated dependencies reported by \depclean at the $i^{th}$ commit on the \pom.
We collect two series for each project, for bloated-direct and bloated-transitive dependencies. 

For each project $p$, we determine the overall trend for the evolution of the number of bloated dependencies: increase, decrease or stable. 
The following function over $\mathcal{B}_p$ shows how we determine the trend for a project:
\[
  f(\mathcal{B}_p) =
  \begin{cases}
    \text{\emph{inc}} & \text{if $slope(lm(\mathcal{B}_p)) > 0 \wedge \exists b_j \in \mathcal{B}_p : b_j < b_{j-1}$} \\
    \text{\emph{dec}} & \text{if $slope(lm(\mathcal{B}_p)) < 0 \wedge \exists b_j \in \mathcal{B}_p : b_j > b_{j-1}$} \\
    \text{\emph{stable}} & \text{if $\forall b_i \in \mathcal{B}_p : b_i = b_{i-1}$} 
  \end{cases}
\]

We notice that several projects do not have a monotonic trend in the number of bloated dependencies (\ie the value increases and decreases at different time intervals). 
To account for projects that have a non-monotonic number of bloated dependencies, we fit a simple linear regression model, denoted as \emph{lm}, and determine the trend of the time series based on the sign of the \emph{slope} of the regression line. 
A project labelled as \emph{inc} is a project for which the sign of the \emph{slope} is positive, \ie, the number of bloated dependencies increase over time.
A project labelled as \emph{dec} is a project for which the sign of the \emph{slope} is negative, \ie, the number of bloated dependencies decreases over time.
If the number of bloated dependencies is the same across all the data points in the time series of a project, we label it as \emph{stable}.

\subsection{Methodology for RQ2}\label{sec:methodology_rq2}

In this research question, we analyze the evolution of the usage status of the \nbDependencies dependencies in our dataset.
Given a dependency $d \in \mathcal{D}$, present in the dependency tree of a project $p$, we collect the status of $d$ at each analyzed commit (see data collection  \autoref{sec:data_collection}). This provides a sequence of usage statuses for $d$ and serves as the basis to determine the occurrence of  transitional patterns between used and bloated statuses.

Let $\mathcal{V}_d$ be a vector representing the history of usage statuses of dependency $d$ across the releases of a project, where each release is ordered by its date.
We label the usage status of a dependency $d$ as \texttt{B} if it is a bloated dependency, or \texttt{U} if it is a used dependency. 

\begin{figure}[t]
	\centering
	\includegraphics[origin=c,width=0.4\textwidth]{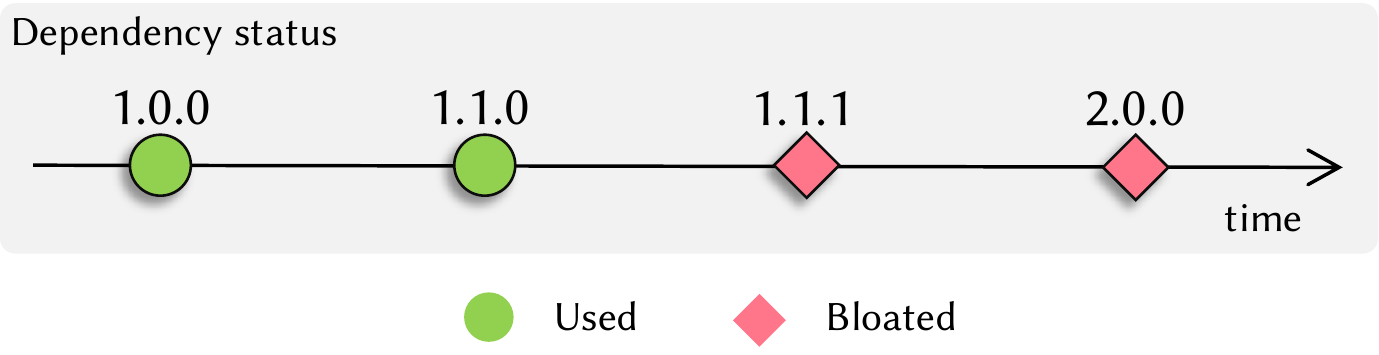}
	\caption{Example of a dependency analyzed over time.  It has a transition of usage status: from used at version 1.1.0 to bloated at  1.1.1 (RQ2). The dependency has two subsequent updates after bloated: at versions 1.1.1 and 2.0.0 (RQ3).} 
	\label{fig:meth_rq2}
	\vspace{-0.4cm}
\end{figure}

\autoref{fig:meth_rq2} illustrates a transition in the usage status of a dependency from used (\texttt{U}) to bloated (\texttt{B}).
In this case, the dependency is identified as used at the two first releases of the project, then it becomes bloated at the third release, and stays as such.
Therefore, the usage pattern for this dependency results in $[\texttt{U},\texttt{U},\texttt{B},\texttt{B}]$.
Since we are interested in analyzing transitional patterns, the consecutive elements of the same category in the vector can be compressed to a single status, \eg, the previous example is represented as $[\texttt{U},\texttt{B}]$.

In this research question, we focus on analysing the occurrence of five transitional patterns: $[\texttt{U}]$, $[\texttt{B}]$, $[\texttt{U,B}]$, $[\texttt{B,U}]$, and \textit{fluctuating}. 
In the cases where the usage status of a dependency  flickers over time, we consider the status of the dependency as \textit{fluctuating}.

\subsection{Methodology for RQ3}\label{sec:methodology_rq3}

We conjecture that developers could save some maintenance efforts in the absence of bloated dependencies.
In this research question, we investigate how many times developers update the version of dependencies that are in fact bloated.
This type of change in the \pom of a project is an unnecessary engineering effort that could be avoided.
We analyze two types of commits: the commits where the developers update the version of the project to a new stable version (\eg, 1.0.0), and the Dependabot commits.
Dependabot\footnote{\url{https://dependabot.com}} is a dependency management bot very active on GitHub. 
It creates pull requests that update the dependencies to remove known vulnerabilities.
Dependabot was launched on May 26, 2017 with support for Ruby and JavaScript, and now it is supporting more than ten languages, including Java since August, 2018.

We analyze Dependabot commits because they only contain edits on the dependency versions in the \pom. 
It provides a clean point of analysis to detect the impact of a dependency update.
And it allows us to study how many bloated dependencies are updated by developers as a result of the suggestion of automatic bots.

\autoref{fig:meth_rq2} illustrates a typical case of a dependency that continues to be updated even after it becomes bloated. 
The dependency is used by the project until version 1.1.0. 
Afterward, the dependency is no longer used, but it is still updated twice,  to version 1.1.1 and then to version 2.0.0.

To answer this research question, we consider the \np{\nbTotalProjectsWithDependabotUpdates} commits in our dataset that perform dependency updates in projects that have at least one Dependabot commit. 
We obtain the number of times a dependency is updated by a developer, by Dependabot, and how many of those updates are performed on bloated dependencies.
For the dependency usage analysis, we tag each dependency as used or bloated. 
We count every time the version of a direct dependency is updated, and we count separately the number of updates applied on bloated dependencies.
In the example of \autoref{fig:meth_rq2}, we count one update on a used dependency (when the used dependency is updated to version 1.1.0), and two updates on a bloated dependency (when the bloated dependency is updated to version 1.1.1 and version 2.0.0).
Using this approach, we can compare the ratio of updates made by developers and by Dependabot. 

\subsection{Methodology for RQ4}\label{sec:methodology_rq4}

In this research question, we investigate the origins of bloated dependencies. 
Each time a bloated dependency appears for the first time in a project's history, we first determine if it was used in the commit that immediately precedes the apparition of bloat.
If the dependency was used in the previous commits, we  determine in which class it was used.
By analyzing a dependency at the time it appears as bloated, we can identify what causes the emergence of bloat.
We have identified four different situations:
\vspace{-.3em}
\begin{enumerate}[leftmargin=0.6cm,noitemsep]
    \item New dependency (\texttt{ND}): The bloated dependency was not present in the previously analyzed commit. It indicates that the dependency was introduced in the project but never used.
    \item Removed code (\texttt{RC}): The bloated dependency was present in the previously analyzed commit and all the classes where the dependency was used are removed.
    \item Updated code (\texttt{UC}): The bloated dependency was present in the previously analyzed commit, yet at least one class  where the dependency was used is still present in this commit. It means that the code has been updated to remove the usage of the dependency but the \pom still contains the dependency.
    \item New version (\texttt{NV}): The bloated dependency was present in the previously analyzed commit and the version of the dependency changed. In the case of transitive dependency, the parent dependency has been updated and the project does not use  the transitive dependency anymore. 
\end{enumerate}
\vspace{-.3em}
For each of the \nbCommits dependency trees, we identify the bloated dependencies.
Then, we check the status of the dependency in the previous commit.
If the dependency is not present in the previous commit, we consider the origin as \texttt{ND}.
Otherwise, we check in the previous commit in which classes the bloated dependency is used.
We then compare those classes with the new commit. 
If all classes are removed, we consider the origin of the bloat as \texttt{RC}.
If at least one of the classes is still present, we consider the origin of the bloat as \texttt{UC}.
Additionally, we compare the version of the bloated dependency with the previous commit. 
If the version  changes, and at least one class is still present, we mark the origin as \texttt{UC} and \texttt{NV}, 
since both reasons could be the origin of the bloat.

\section{Results}

In this section, we answer the four RQs presented in \autoref{sec:rqs}.

\subsection{RQ1. Bloat Trend}

In this research question, we analyze the evolution of the number of bloated dependencies over time. 
We hypothesize that this number tends to grow.
Following the protocol described in \autoref{sec:data_collection}, we analyze the usage status of each dependencies in \nbCommits dependency trees along the history of \nbProjects projects, as reported by \depclean.
We assign bloat trend labels to each project, according to the three categories defined in \autoref{sec:methodology_rq1}.

\begin{figure}[htb]
	\centering
	\includegraphics[origin=c,width=0.43\textwidth]{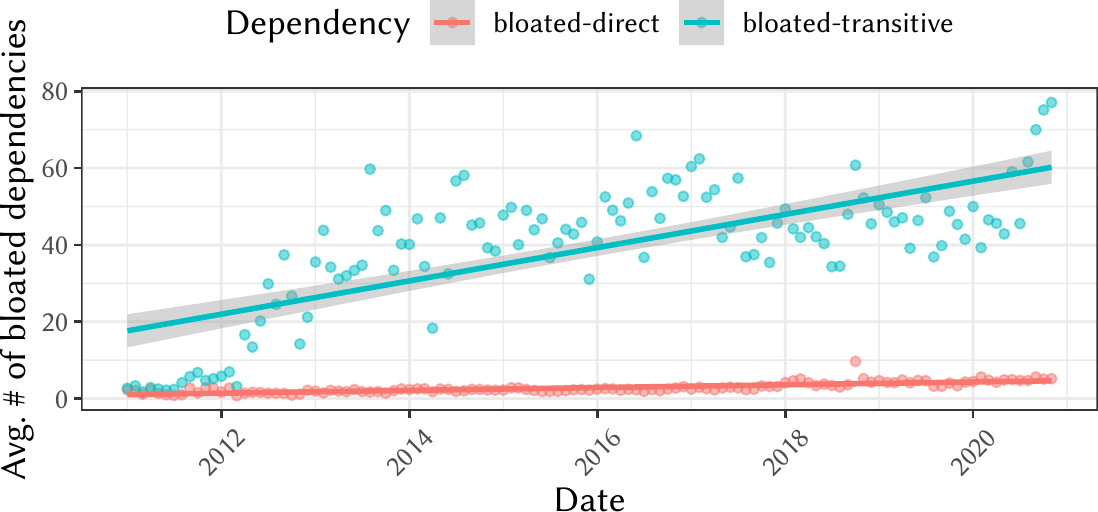}
	\caption{Trend of the average number of bloated-direct and bloated-transitive dependencies per month.}
	\label{fig:trend_bloat_average}
	\vspace{-0.25cm}
\end{figure}

\autoref{fig:trend_bloat_average} shows the monthly evolution trend of the number of bloated-direct and bloated-transitive dependencies, from January 2011 to November 2020.
The y-axis is the average number of bloated dependencies of the \nbProjects projects.
Each data point represents the average of bloat measured each month.
The lines represent linear regression functions, fitted to show the trend of bloated-direct and bloated-transitive dependencies, at a $95\%$ confidence interval.

\FPset{nbTransitiveFirst}{1,695}
\FPset{nbTransitiveLater}{286,228}

We observe that bloated-transitive dependencies have a clear tendency to grow over time, whereas bloated-direct dependencies grow at significantly lower pace.
For example, the number of bloated-transitive dependencies in 2011 was \nbTransitiveFirst, and by the end 2020 this number grew up to \nbTransitiveLater ~(increase > $250\times$).
The bloat is more pervasive and variable (SD = 17.2) among transitive dependencies, representing a larger share in comparison with direct dependencies that are less numerous and less variable (SD = 1.3).
We conclude that, overall, the amount of bloat increases, being more notable for transitive dependencies.



\FPset{increasingProject}{245}
\FPset{decreasingProject}{106}
\FPset{stableProject}{84}

\FPset{increasingProjectTransitive}{286}
\FPset{decreasingProjectTransitive}{113}
\FPset{stableProjectTransitive}{36}

\begin{figure}[htb]
	\centering
	\includegraphics[origin=c,width=0.45\textwidth]{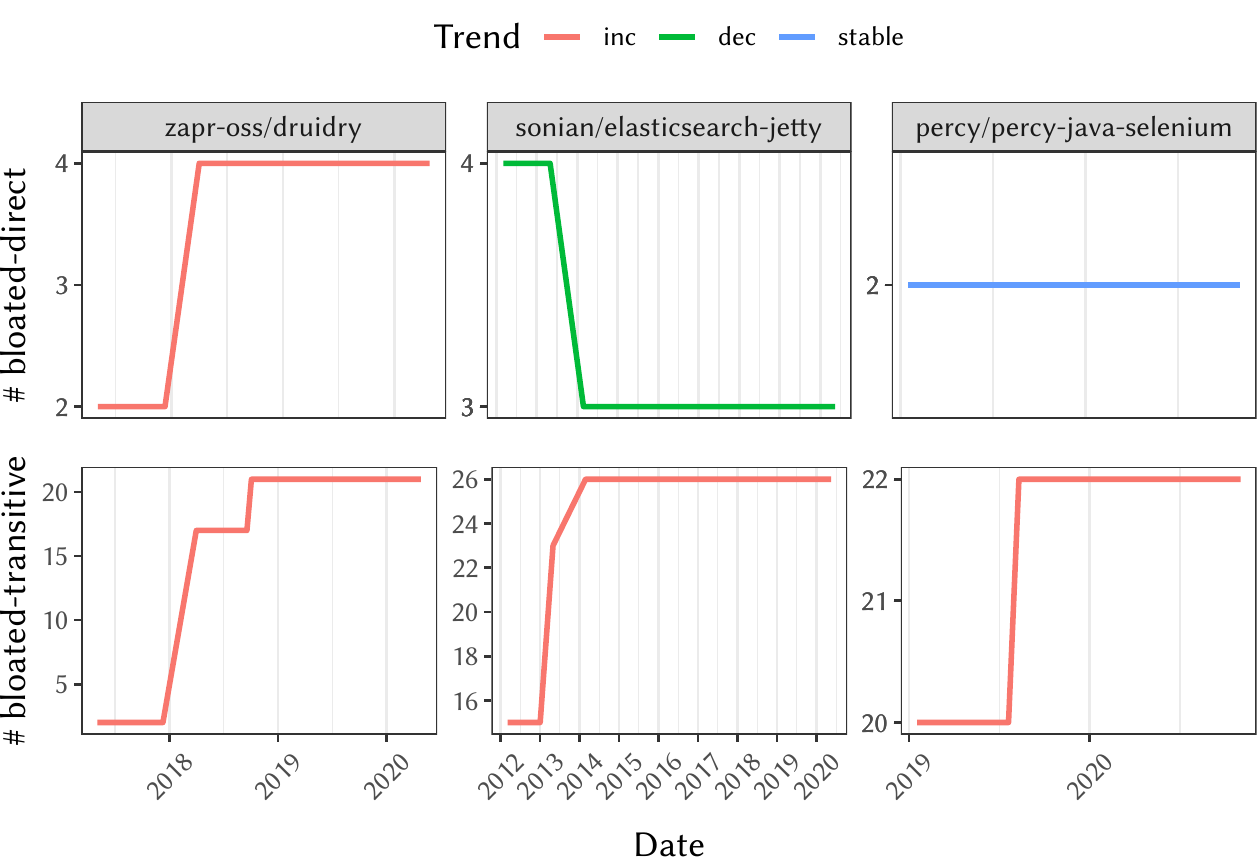}
	\caption{Example of projects in the three classes of bloat trend defined in \autoref{sec:methodology_rq1}.}
	\label{fig:lineplot_projects}
	\vspace{-0.25cm}
\end{figure}

\autoref{fig:trend_bloat_average} shows an overall growing trend for the number of bloated dependencies. Now, we look in more details at each project separately. 
We count the number of projects that have different trend of bloated dependencies.
\autoref{fig:lineplot_projects} shows examples of time series of projects in our dataset for which the bloated-direct dependencies are labelled according to each category (increasing, decreasing, and stable).
The name of the projects correspond to the <\textit{user}>/<\textit{repository}> on GitHub.
The x-axis is the date of the analyzed commits.
The y-axis represents the number of bloated dependencies detected.
For instance, the time series of the project \href{https://github.com/zapr-oss/druidry}{zapr-oss/druidry} has a total of \np{51} commits on the \pom (\ie, data points in the time series), and it is labelled as \emph{inc} \wrt to both the direct and transitive dependencies because both series tend to continuously increase over time.

\begin{figure}[b]
	\centering
	\includegraphics[origin=c,width=0.475\textwidth]{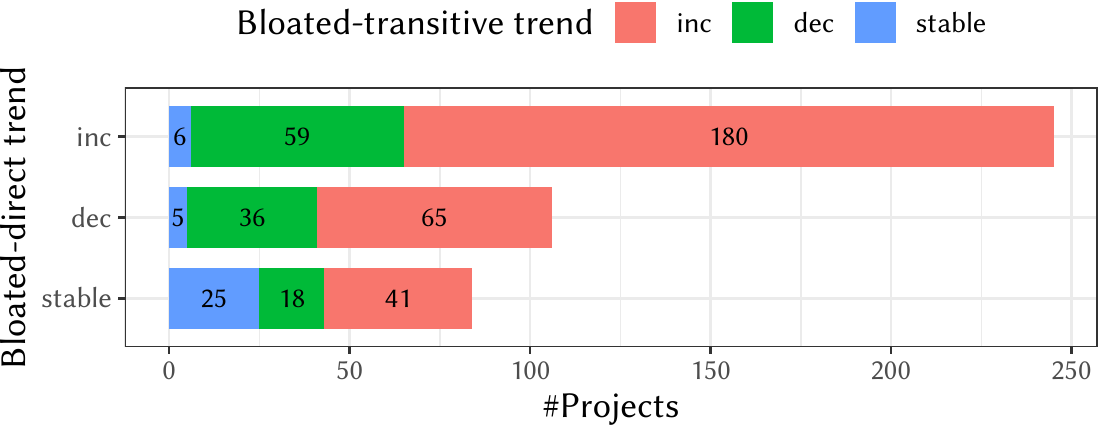}
	\caption{Distribution of the number of projects with increasing, decreasing, and stable trend of bloated-direct and bloated-transitive dependencies.}
	\label{fig:barplot_trend}
\end{figure}

\autoref{fig:barplot_trend} shows the distribution of the trend of bloated-direct and bloated-transitive dependencies.  
The x-axis indicates the number of projects with bloated-direct dependencies in each specific evolution trend, given on the y-axis. Each bar in the plot is partitioned in three parts that correspond to the share of projects with a given trend for the number of bloated-transitive dependencies.
For example, the top bar of \autoref{fig:barplot_trend} shows (i) that the number of bloated-direct dependencies tends to increase for \percent[p]{\increasingProject}{\nbProjectsNum} projects; and (ii) among these \np{\increasingProject} projects, \np{180} also have a number of bloated-transitive dependencies that tends to increases, \np{59} of these projects have a decreasing number of bloated-transitive dependencies and \np{6} projects have a stable number of bloated-transitive dependencies.
The bar in the middle of the figure indicates that the number of bloated-direct dependencies tends to decrease for \percent[p]{\decreasingProject}{\nbProjectsNum} projects and the bottom bar shows that this type of bloat is stable for \percent[p]{\stableProject}{\nbProjectsNum} projects because no new bloated dependencies are introduced in the \pom.

Looking at the partitions of each bar in \autoref{fig:barplot_trend}, we first observe that whatever the trend for the number of bloated-direct dependencies, the number of bloated-transitive dependencies can evolve in any way. Yet, the majority of projects have an increasing number of bloated dependencies among their transitive dependencies.
In total, \percent[p]{\increasingProjectTransitive}{\nbProjectsNum} projects have an increasing number of bloated-transitive, whereas for \percent[p]{\decreasingProjectTransitive}{\nbProjectsNum} projects this number decreases.
The number of projects with stable transitive-dependencies, \percent[p]{\stableProjectTransitive}{\nbProjectsNum}, is relatively low.

Interestingly, from the \np{\stableProject} projects with a stable number of bloated-direct dependencies, \percent[p]{41}{\stableProject} of the bloated-transitive dependencies increase and \percent[p]{18}{\stableProject} decreases (\eg, as in the project \href{https://github.com/percy-java-selenium}{percy/percy-java-selenium} in \autoref{fig:lineplot_projects}).
This result indicates that the usage status of dependencies change regardless of the modification of the \pom.
The transition from used to bloated in transitive dependencies becomes unnoticed.
In other words, even if developers update only the version of direct dependencies, without doing anything else, then the bloat grows naturally due to the inflation of the rest of the dependency tree.
It happens, for example, in the project \href{https://github.com/jpmml/jpmml-sparkml/}{jpmml/jpmml-sparkml} when a developer updates \texttt{spark-mllib\_2.11} from version 2.0.0 to 2.2.0, introducing \np{133} new transitives dependencies.

On the other hand, we observe that for \percent[p]{65}{\decreasingProject} out of  the \np{\decreasingProject} projects with a decreasing number of bloated-direct dependencies,  the number of bloated-transitive increases. 
It indicates that even in projects for which direct dependencies decreases, the number of bloated-transitive dependencies can increase and eventually lead to a global growth of bloated dependencies for the project.

\vspace{+0.30cm}
\begin{mdframed}[style=mpdframe]
\textbf{Answer to RQ1:} 
The number of bloated-direct dependencies and bloated-transitive dependencies increases over time for \ShowPercentage{\increasingProject}{\nbProjectsNum} and \ShowPercentage{286}{\nbProjectsNum} of the projects, respectively.
This result suggests that bloated dependencies tend to naturally emerge and grow through software evolution and maintenance.
\end{mdframed}
\vspace{-0.15cm}

\subsection{RQ2. Usage Patterns}

This research question addresses an essential concern when developers think about removing bloat: is a piece of software identified as bloat at one point in time prone to usage in future revisions? We answer this question through a post-mortem analysis of the transitioning in the usage status of dependencies across the evolution of the studied projects.
Our hypothesis is that dependencies do not change their usage status very frequently, \ie, a dependency that is used in one commit is used in future commits, and similarly for bloated dependencies.
If our hypothesis holds, then it indicates that developers can be more confident when removing bloated dependencies.\looseness=-1

We analyzed the five usage patterns described in \autoref{sec:methodology_rq2}. \autoref{fig:tterns} shows one concrete example for each pattern. The examples are taken from our dataset and the patterns are illustrated on the period January 2017 to December 2020.
The y-axis shows the name of the direct dependency, with the pattern in square brackets.
For example, we analyze the usage status of the direct dependency \texttt{h2} in the project \href{https://github.com/dieselpoint/norm}{dieselpoint/norm}, from May 2018 to October 2020. 
As we can observe, this dependency was always reported as bloated.
On the other hand, the dependency \texttt{json} in project 
\href{https://github.com/PAXSTORE/paxstore-openapi-java-sdk}{PAXSTORE/paxstore-openapi-java-sdk} was reported as bloated in first four analyzed commits, September 2018 to November 2019, and then it was used in all the subsequent releases of the project.\looseness=-1

\begin{figure}[htb]
	\centering
	\includegraphics[origin=c,width=0.475\textwidth]{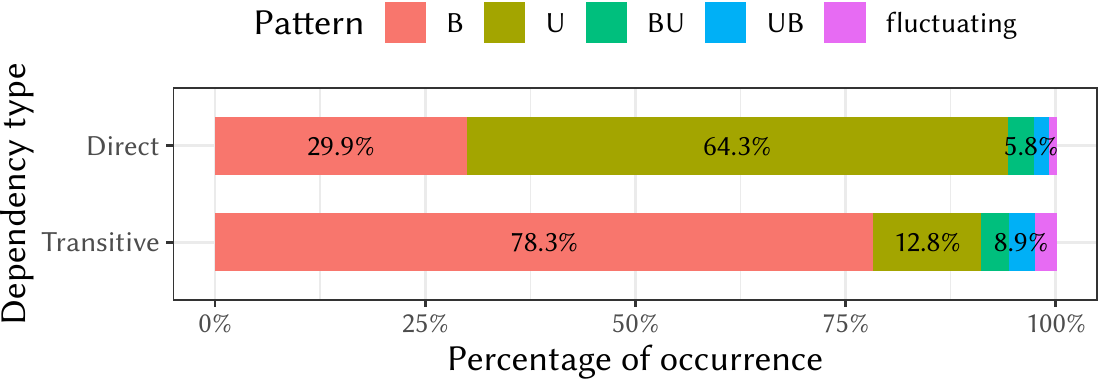}
	\caption{Percentage of occurrence of usage patterns of bloated-direct and bloated-transitive dependencies.}
	\label{fig:barplot_patterns}
\end{figure}

\autoref{fig:barplot_patterns} shows the distribution of the five transitional usage patterns among the \np{\NbOriginBloatDirectDependencyNum} direct  and  \np{\NbOriginBloatTransitiveDependencyNum} transitive dependencies in our dataset.
The x-axis represents the percentage of occurrence of each pattern with respect to the total.
The top bar of the plot indicates that \np[\%]{64.3} of the direct dependencies are used through their whole lifespan, whereas \np[\%]{29.9} are always bloated.
This means that \np[\%]{94.2} of direct dependencies never change their status through the evolution of the software projects.
This also means that most bloated-direct dependencies are bloated by the time they are added in the dependency tree and are likely to remain bloated forever. 
We conjecture that this happens as a side effect of some development practices, such as copy-pasting of \pom files, the use of Maven Archetypes, or the deliberate addition of dependencies when setting up the development environment.

The bottom bar of the plot shows a similar stability for the status of transitive dependencies: \np{91.1}\% of transitive dependencies do not change their usage status over their lifespan. A key difference here is that most of the dependencies are always bloated: \np{78.3}\% of the transitive dependencies are bloated from the start, whereas \np{12.8}\% are always used.
We hypothesize that most transitive dependencies are unnoticed by the developers. Consequently, they are not managed and stay in the dependency tree for no reason in most cases.\looseness=-1

\vspace{-0.15cm}
\begin{figure}[htb]
	\centering
	\includegraphics[origin=c,width=0.43\textwidth]{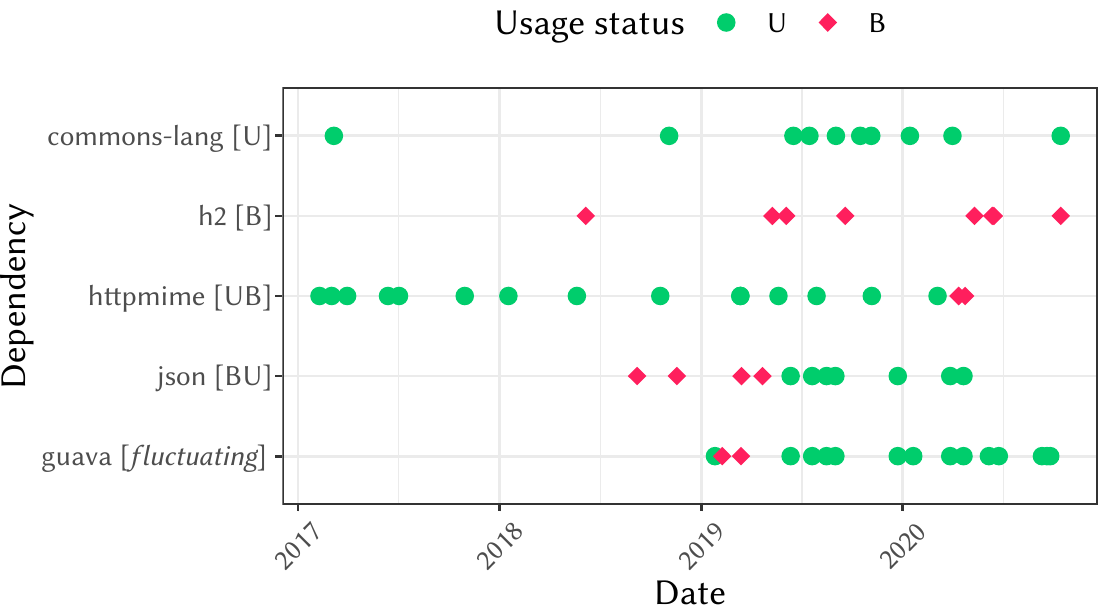}
	\caption{Example of direct dependencies with distinct usage patterns. Each dependency belongs to a different project, the status of the dependency is analyzed at each commit that changes the \pom of the project.}
	\label{fig:example_patterns}
	\vspace{-0.25cm}
\end{figure}

A key motivation for this research question is to determine whether a dependency identified as bloated is likely to stay bloated.
We compute the percentage of dependencies bloated from the start (\texttt{B}) or that remain bloated after being used (\texttt{UB}), with respect to the total number of dependencies that are bloated at some point in the future, \ie, (\texttt{B}+\texttt{UB})/\texttt{(B}+\texttt{UB}+\texttt{BU}+\textit{fluctuating}).
We find that \ProbBloatedDirect of bloated-direct dependencies and \ProbBloatedTransitive of bloated-transitive dependencies remain bloated over time. 

\begin{mdframed}[style=mpdframe]
\textbf{Answer to RQ2:} 
The usage status is mostly constant over time: \np{94.2}\% of the direct and of \np{91.1}\% of the transitive dependencies do not change status through their lifespan \ie they are either always used or always bloated.
When a dependency is detected as bloated, it stays bloated in \ProbBloatedDirect of the cases if it is direct, and \ProbBloatedTransitive if is transitive.
This indicates that developers can confidently take a debloating action when detecting bloated dependencies.
\end{mdframed}

\subsection{RQ3. Unnecessary Updates}

In this research question, we investigate how the update of dependencies, a regular maintenance practice for all software projects, more and more encouraged by automatic bots, relates to bloated dependencies.
We hypothesize that developers invest some effort in updating some of these dependencies, while this is not required.
To verify this hypothesis, we count how many times bloated-direct dependencies are updated in the \pom and compare it to the number of updates of used-direct dependencies. 
The methodology for this count is described in \autoref{sec:methodology_rq3}.
We analyze separately the updates  performed manually by developers and the updates suggested by Dependabot that are eventually accepted by a developer. 

\begin{figure}[t]
	\centering
	\input{figs/chart_bloat_dependabot_update}
	\caption{Number of updates made by Dependabot on direct dependencies in projects that use Dependabot.}
	\label{fig:barplot_dependabot_updates}
\end{figure}
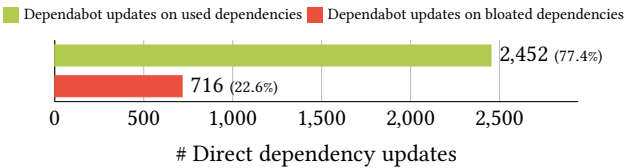

Figures \ref{fig:barplot_updates} and \ref{fig:barplot_dependabot_updates} present our main results for this research question. 
Those plots present the number of dependency updates on direct dependencies made by developers and by Dependabot respectively.
We focus on the \np{143} projects in our dataset that have at least one Dependabot commit.
All the projects do not use Dependabot since its Java support is relatively recent (August 2018). 
The total number of updates on direct dependencies in these projects is \np{\nbTotalProjectsWithDependabotUpdates}, of which \np{\nbDepDevUpdatesNum} ~have been performed by humans and \np{\nbDependabotUpdatesNum} ~have been suggested by Dependabot.

\autoref{fig:barplot_updates} shows that, over a total of \np{\nbDepDevUpdatesNum} updates on direct dependencies made by developers,
\percent[p]{\nbDepUsedUpdateNum}{\nbDepDevUpdatesNum} are preformed on used dependencies, and \percent[p]{\nbDepBloatUpdateNum}{\nbDepDevUpdatesNum} are made on bloated dependencies.
These \np{\nbDepBloatUpdateNum} unnecessary updates represent a significant effort, as updating dependencies is a non trivial maintenance task \cite{Kula2018}.
\autoref{fig:barplot_dependabot_updates} shows the number of updates on direct dependencies made by accepting a suggestion from Dependabot. 
From \autoref{fig:barplot_dependabot_updates}, \percent[p]{\nbUsedDependabotUpdateNum}{\nbDependabotUpdatesNum} of Dependabot updates are performed on non-bloated dependencies and \percent[p]{\nbBloatDependabotUpdateNum}{\nbDependabotUpdatesNum} on bloated dependencies.
Overall, we observe that developers perform a significantly larger number of dependency updates than Dependabot. Yet, the most interesting fact is that  developers and Dependabot perform the same ratio of updates on bloated dependencies, \np[\%]{22.0} and \np[\%]{22.6} respectively.


\begin{figure}[t]
	\centering
	\input{figs/chart_bloat_update}
	\caption{Number of updates made by developers on direct dependencies in projects that use Dependabot.}
	\label{fig:barplot_updates}
\end{figure}
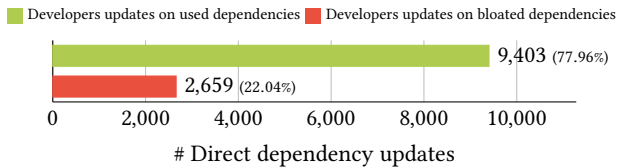

The consequences of updating a bloated dependency are not only about the time and effort wasted by the developer. We have observed that a possible side-effect of these unnecessary updates is the increase of the total number of bloated dependencies in the project.
In RQ1, we showed that the number of bloated dependencies increases over time, with a strong trend for transitive dependencies. 
In fact, a portion of this increasing transitive bloat is introduced through the update of direct dependencies, \ie, the new version has more dependencies.
Note that this scenario can happen even when updating a bloated-direct dependency.
We have observed this phenomenon in our dataset. 
The \np{\nbTotalBloatUpdatesNum} updates on bloated-direct dependencies have introduced \np{\nbBloatIntroducedByUpdateNum} new bloated-transitive dependencies.

\vspace{+0.30cm}
\begin{mdframed}[style=mpdframe]
\textbf{Answer to RQ3:} 
\np[\%]{22.0} of developer updates and \np[\%]{22.6}
of Dependabot accepted updates are performed on  bloated-direct dependencies, which represents a total of   \np{\nbTotalBloatUpdatesNum} updates over \np{143} projects. 
This is novel evidence that software bloat artificially increases maintenance effort and that dependency bots need to be improved to detect bloated dependencies.
\end{mdframed}

\subsection{RQ4. Bloat Origin}

In this research question, we investigate what type of maintenance activity is at the origin of bloat emergence. 
In other words, we perform an in-depth analysis of the usage patterns \texttt{B} and \texttt{UB} presented in RQ2 by categorizing the origin of the bloat in  four possible activities: new dependency (\texttt{ND}), removed code (\texttt{RC}), updated code (\texttt{UC}), and new version (\texttt{NV}) as described in \autoref{sec:methodology_rq4}.
\autoref{tab:origin_bloat} summarizes the number of occurrences of activities that introduce bloat for direct and transitive dependencies.
In total, we analyze the \np{\NbOriginBloatDependencyNum}  dependencies that become bloated at some point in time (\np{\NbOriginBloatDirectDependencyNum} directs, \np{\NbOriginBloatTransitiveDependencyNum} transitives) and determine in what condition they become bloated. This corresponds to  \np{\totalOriginDirect} and \np{\totalOriginTransitive} transitions to bloat, on direct and transitive dependencies respectively.

We observe that the primary origin of bloat is the addition of new dependencies \texttt{ND}, with \percent[p]{\originDirectBloatNewDepNum}{\totalOriginDirect} such additions that lead to more bloated-direct  dependencies and \percent[p]{\originTransitiveBloatNewDepNum}{\totalOriginTransitive} new dependencies that introduce more bloated-transitive dependencies.
This result confirms our findings in RQ2, where we observed that the status of most dependencies does not change  over time, which hinted to the fact that bloated dependencies are bloated as soon as they appear in the dependency tree. 
Additionally, the larger number of \texttt{ND} that grow the number bloated-transitive dependencies  is consistent with the results of RQ1, where we showed a larger increase of bloated-transitive dependencies than bloated-direct ones. This new result consolidates the finding with the root cause of the transitive bloat.
The second most frequent origin of bloated dependencies is different for direct and transitive dependencies. The action of removing code
\texttt{RC} is the second most frequent cause of the emergence of bloated-direct dependencies, with $8,8\%$ of the cases. Updating code \texttt{UC} is the second most important root cause for bloated-transitive dependencies. 
While these two actions are similar in nature (evolve the code base), we did not find a clear explanation for the difference between the types of bloated dependencies.
Updating to a new version of a dependency \texttt{NV} is the least frequent cause of bloat emergence. The rarity of this cause is explained by the fact that it can only happen in very specific conditions, when the new version of the dependency changes drastically. 

\begin{table}[t]
\small
    \caption{Number of occurrence for each origin of bloat. The occurrences are separated between the new bloated-direct and transitive dependencies.}
    \vspace{-0.25cm}
    \label{tab:origin_bloat}
    \centering
    \begin{tabular}{@{}lrr@{}}
    \toprule
Origin                    & Bloated-direct & Bloated-transitive  \\
\midrule
New dependency (\texttt{ND}) & \percent{\originDirectBloatNewDepNum}{\totalOriginDirect} & \percent{\originTransitiveBloatNewDepNum}{\totalOriginTransitive} \\
Removed code  (\texttt{RC})  & \percent{\originDirectBloatRemoveNum}{\totalOriginDirect} & \percent{\originTransitiveBloatRemoveNum}{\totalOriginTransitive} \\
Updated code (\texttt{UC})   & \percent{\originDirectBloatUpdateNum}{\totalOriginDirect} & \percent{\originTransitiveBloatUpdateNum}{\totalOriginTransitive} \\
New version (\texttt{NV})  & \percent{\originDirectBloatNewVersionNum}{\totalOriginDirect}  & \percent{\originTransitiveBloatNewVersionNum}{\totalOriginTransitive} \\
\bottomrule
    \end{tabular}
\vspace{-0.2cm}
\end{table}

We now illustrate the different situations of bloat introduction with real-world case studies observed in our dataset.
The most frequent cause of bloat introduction is a new transitive dependency in the dependency tree  (\texttt{ND}), which is never used.
For example, this happens in the project \href{https://github.com/couchbase/couchbase-java-client}{couchbase/couchbase-java-client} at the commit \href{https://github.com/couchbase/couchbase-java-client/commit/47ac44fa1aa2eb0271b4823fbe9f7d6a69818a95}{47ac44}, where the dependency 
\texttt{jackson-\allowbreak{}databind}, 
which is induced transitively when 
\texttt{encryption\allowbreak{}:1.0.0} has been added to the \pom.
\texttt{jackson-\allowbreak{}databind} is used in the class 
\texttt{Hashicorp\allowbreak{}VaultKey\allowbreak{}StoreProvider} which is never used by the \href{https://github.com/couchbase/couchbase-java-client}{couchbase/couchbase-java-client} and, therefore,
\texttt{jackson-\allowbreak{}databind}
is a bloated-transitive dependency in this project.

This case occurs with direct dependencies as well.
For example,  the direct dependency 
\texttt{jackson-core} is added as a direct dependency in the \pom of project \href{https://github.com/jenkinsci/elasticbox-plugin}{jenkinsci/elasticbox-plugin}, at commit \href{https://github.com/jenkinsci/elasticbox-plugin/commit/0083584a760c6800ea63bae8283761cadf2ef448}{008358}. Yet, the dependency is never used in the code of the project.
One year and 4 months later a pull-request, \href{https://github.com/jenkinsci/elasticbox-plugin/pull/41}{\#41}, fixes the bloat issue by removing the dependency directly. 
However, at the time of writing this paper, the pull-request has not been merged.

Projects are evolving, adding and removing code is part of the life cycle of a project. A consequence of code removal can be to eliminate the need for a dependency.
Yet, developers currently have no tool support to determine that a dependency can also be removed as part of their maintenance activities. Consequently, the dependency is likely to become bloated (\texttt{RC}).
For example, we observed that scenario happens in the project \href{https://github.com/apache/commons-lang}{apache/commons-lang}. 
The commit \href{https://github.com/apache/commons-lang/commit/def3c4672b1f99b35ec1fb1aaa316e6043463d73}{def3c4} introduces the dependency 
\texttt{bcel}, which contains annotations to document thread safety.
However, the commit \href{https://github.com/apache/commons-lang/commit/796b051f28ca96f1dbbd5dfe4b5cae5579d0d14e}{796b05} removes all classes where these annotations were used.
According to the commit, more discussions were needed to design the annotation, and the maintainers reverted partially the changes to release a new version.
A developer removed the bloated dependency after five months (see commit \href{https://github.com/apache/commons-lang/commit/66226ec1c2ff33e138189463001c649dbb404f56}{66226e}).

A similar scenario occurs when developers  update classes (\texttt{UC}).
For example, the commit \href{https://github.com/jenkinsci/remoting/commit/62aad35f64c46d0c41a5d464ccba8c78b784f9d7}{62aad3} introduces   the annotation \texttt{Ignore\allowbreak{}JRERequirement} on a method in the project \href{https://github.com/jenkinsci/remoting/}{jenkinsci/remoting}. 
However, this method is updated and deprecated in the commit \href{https://github.com/jenkinsci/remoting/commit/49c67eef8616c7bc3588263d4ecc2dbcb51d5bb8}{49c67e}.
The annotation \texttt{Ignore\allowbreak{}JRERequirement} is removed and the dependency
\texttt{animal-\allowbreak{}sniffer-\allowbreak{}annotation} 
became bloated.

The project \href{https://github.com/apache/commons-dbcp/}{apache/commons-dbcp} contains an interesting case of bloat introduced when a dependency is updated (\texttt{NV}).
In the commit \href{https://github.com/apache/commons-dbcp/commit/3550adaf1c87af8ec9b80830f9d9e9d2a1e19691}{3550ad}, the direct dependency 
\texttt{geronimo-\allowbreak{}transaction} 
is detected as bloated.
However, this dependency was not bloated in the previous commit \href{github.com/apache/commons-dbcp/commit/d7aa662fbbb99e536ae28c47d0c4e1d51e39d5b9}{d7aa66}, when the project was using the version \texttt{1.2-beta} of 
\texttt{geronimo-\allowbreak{}transaction}. The dependency  was updated to  version \texttt{2.2.1} with commit \href{github.com/apache/commons-dbcp/commit/3550adaf1c87af8ec9b80830f9d9e9d2a1e19691}{3550ad}. 
This new version brought major changes in the dependency and, in \texttt{2.2.1}, all the classes used by the project had been move in a transitive dependency of 
\texttt{geronimo-\allowbreak{}transaction}. 
Therefore, the direct dependency towards
\texttt{geronimo-\allowbreak{}transaction}
became bloated.

\vspace{+0.15cm}
\begin{mdframed}[style=mpdframe]
\textbf{Answer to RQ4:} The addition of new dependencies is the root cause of the emergence of 
\ShowPercentage{\originDirectBloatNewDepNum}{\totalOriginDirect} of the bloated-direct dependencies. 
Still, \ShowPercentage{347}{\totalOriginDirect} of the cases appear after code updates or removals.
This indicates that new dependencies should be carefully reviewed the first time they are added, and we recommend developers to check the usage status of dependencies when removing code.\looseness=-1
\end{mdframed}
\vspace{-0.2cm}

\section{Implications}\label{sec:implications}


Our findings provide practical, empirically justified implications for improving dependency management \cite{Cox2019,Gustavsson20}.
The results of RQ1 and RQ2 show that bloated dependencies are likely to remain bloated in the future.
This is empirical evidence that can motivate developers and increase their confidence when they are faced with the opportunity to remove bloated dependencies. Motivation comes from the benefits associated with reducing the number of dependencies of the project and hence reduce associated maintenance activities. Confidence comes with the strong likelihood that the dependency that is removed will not be necessary in the future.

Our results show that there exist many practical difficulties related to the way of handling software dependencies.
This can raise the awareness of developers about the importance of understanding what dependencies are more likely to become bloated, and how their projects can reduce the size of dependency trees without breaking the build.
In particular, the use of tools, such as \depclean, to automatically detect and suggest changes in the build files can contribute to a better awareness of developers about the state of their dependencies. 
For example, we recommend to include a bloat analysis before release to ensure that no bloat is shipped and deployed. 
This is for reducing the size of the released binary and for all projects that depend on it, \ie, the number of transitive dependencies will be reduced.\looseness=-1

In RQ3, we present original results of the negative impact of bloated dependencies on the maintenance of the projects. 
In particular, we shed a new light on the limitations of dependency bots, such as Dependabot, and provide evidence that developers accept bots' suggestions when updating dependencies without checking if the dependency is actually used.
Bot creators should consider improving their tools to automatically detect bloat and suggest the removal of unused dependencies.
On the same line, compilers and IDEs should also warn developers when dependencies are not used anymore and when a dependency is introduced without encountering its counterpart usage on code.

Our dataset and our case studies on the origin of bloat provide valuable references for the rapid identification of practices that result in dependency bloat.
Those references can be used to build dedicated bots that ask for additional checks, e.g. when a new dependency appears in the dependency tree, or to establish guidelines for developers when they maintain \pom files. 

\section{Threats to validity}


\textit{\textbf{Internal Validity.}}
The first internal threat relates to the detection of bloated dependencies.
The results of our study are tied to the accuracy of \depclean to find bloated dependencies in Maven projects.
This tool is based on advanced static analysis. 
Therefore, some usages that rely on Java dynamic features might be missed, reporting some used dependencies as bloated.
For example, \texttt{lombok} is a Java library that relies on annotations to manipulate the bytecode at compilation time, adding boilerplate code constructs such as getters and setters.
This mechanism makes the dependency to be flagged as bloated by \depclean, since no reference to this dependency remains in the bytecode of the compiled project.
Nevertheless, we consider that \depclean is a solid tool, evaluated on millions of dependencies, used in industry, and developers have removed hundreds of bloated dependencies thanks to its analysis~\cite{Valero2020}.
The second threat relates to the representativeness of the data and the analysis performed.
We mitigate those threats by collecting a large dataset of projects from multiple domains and  released across several years.
This allows us to draw general conclusions about the evolutionary trend of bloated dependencies, regardless of the existence of some false positives, which are known to be hard to detect using static analysis~\cite{Livshits2015,Landman2017,Reif2018}.

\textit{\textbf{External Validity.}} 
When conducting this study, we focus on bloated Java dependencies in projects that build with Maven.
As explained in \autoref{sec:data_collection}, the analysis of the dependency trees of \nbProjects projects requires compiling and analyzing the bytecode at different time periods.
There were cases where the compilation failed for several reasons, making it difficult to obtain the complete history of dependency changes.
As we consider a large number of open-source projects, we believe our results are generalizable in this specific domain. 
Meanwhile, additional studies with proprietary projects or other programming languages should be considered to consolidate these very first result about software bloat evolution.

\textit{\textbf{Construct Validity.}}
This threat is related to the rationality of the questions asked.
We investigate the evolution of bloated dependencies over time.
To achieve this goal, we focus on four aspects: bloat trend, usage patterns, unnecessary updates, and bloat origins.
We believe that these are rational questions that provide unique and novel insights for researchers and developers.

\section{Related work} 

\textit{\textbf{Software Bloat.}}
Previous research on software bloat has mainly focused on reducing C/C++ binaries to mitigate the security risks associated with unnecessary code~\cite{Sharif2018, Qian2019, Quach2018}.
Holzmann~\cite{Holzmann2015} reports the historical growth in the size of the \texttt{true} command in Unix systems.
Similarly, we observed that the number of bloated dependencies tends to grow over time, whether or not there is a need for it.
In the last years, there is a recent resurgence of interest in debloating Java bytecode~\cite{Jiang2016, Bruce2020,Macias2020,Valero2020traces,Valero2020}.
These tools remove Java bytecode using static and dynamic analysis. 
In contrast, our study focuses on the evolution and the emergence of bloat in Java projects, while spotting some of the current research gaps and tooling for effective dependency management.
Other studies have focused on eliminating bloat in source code~\cite{Vazquez2019}, binary shared libraries~\cite{Agadakos2020}, highly configurable programs~\cite{Koo2019}, or containers~\cite{Rastogi2017}.
Other works have focused on improving the debloat process through various optimizations techniques~\cite{Xin2020, Azad2019,Sun2018,AntignacSS17,HagueLH19}.
As far as we know, we are the first to conduct a longitudinal study to analyze software bloat.

\textit{\textbf{Bloated Dependencies.}}
Our work follows up on our previous study of bloated dependencies ~\cite{Valero2020}. 
Our quantitative and qualitative study of bloated dependencies in the Maven ecosystem, revealed the importance of the phenomenon in Maven Central. Our interactions with software developers showed that removing bloated dependencies is perceived as a valuable contribution.
Here we extend this previous study in two ways.
First, we perform a study of bloated dependencies with distinct study subjects on a chronological basis. 
This brings novel insights on the evolution of bloated dependencies through the history of software projects, corroborating the importance of maintaining \pom files.
We bring novel evidence in favor of removing bloated dependencies.
Second, we perform a unique study on the interaction between maintenance activities and the emergence of bloat. These new results contribute to understanding the origin of bloat as well as estimating the maintenance effort unnecessarily invested when performing dependency updates.

\textit{\textbf{Software Bots.}}
Erlenhov \etal~\cite{Erlenhov20} perform an empirical study about the interaction between practitioners and software bots. They found that there is currently a lack of general-purpose smart bots that go beyond simple automation tools, such as dependency version updating. This is in line with our results, as we have seen that dependency bots do not perform advanced dependency usage analysis, sending unnecessary warnings that could be avoided. 
Wessel \etal~\cite{wessel2020inconvenient} rise attention on the inconvenient side of software bots.
They present empirical evidence that pull requests made by bots are, in some cases, perceived as disruptive and unwelcoming by developers.
Thus, motivating our work on reducing the number of warnings caused by bots dedicated to automatically update dependencies.
\vspace{-0.15cm}



\section{Conclusion}
This paper presented the first large-scale longitudinal study about the evolution of software bloat, with a focus on bloated Java dependencies. 
We collected a unique dataset of \nbCommits dependency tree versions, tagged with usage dependency status, from \nbProjects Java projects hosted on GitHub. 
Through the analysis of \nbDependencies dependencies, we provided evidence about an essential phenomenon: \ProbBloatedDirect of the dependencies that become bloated over evolution stay bloated over time.
As a consequence, developers spend significant time updating dependencies that are actually bloated. We find that \np[\%]{22} of dependency updates are made on bloated dependencies. These updates include a significant number of updates suggested by Dependabot.
We also demonstrate that bloated dependencies are primarily originated from the addition of new dependencies that are never used, rather than from code changes.

Our work paves the way to better understand the importance of debloating tools, such as \depclean, to handle the increasing phenomenon of software bloat.
In particular, evidence that bloated code stays bloated is important for developers who need to decide if they should remove code. 
Our novel findings about the role of Dependabot on the unnecessary maintenance effort provide concrete insights to improve the suggestions that this single bot shares with developers.



\begin{acks}
This work is partially supported by the Wallenberg AI, Autonomous Systems, and Software Program (WASP) funded by Knut and Alice Wallenberg Foundation and by
the TrustFull project funded by the Swedish Foundation for
Strategic Research.
\end{acks}

%% file: tables/ts_descriptive.tex
\begin{table}[t]
\footnotesize
\caption{Descriptive statistics of the dependencies in the \nbProjects  analyzed projects.}
\centering
\label{ts_descriptive}
\begin{tabular}{@{}l|c|c|c|c|c|c@{}}
       & Min & 1st Qu. & Median & Avg.  & 3rd Qu. & Max   \\ 
\hline
\# Months & \np{5} & \np{48.5} & \np{75.5} & \np{81.01} & \np{109.5} & \np{235} \\
\# Analyzed commits & \np{2} & \np{41.0} & \np{58.0} & \np{73.51} & \np{79.0} & \np{819} \\
\# Direct initial & \np{0} & \np{3.0} & \np{5.0} & \np{8.28} & \np{10.0} & \np{120} \\
\# Transitive initial & \np{0} & \np{2.0} & \np{10.5} & \np{46.77} & \np{56.0} & \np{300} \\
\# Direct final & \np{0} & \np{5.0} & \np{10.0} & \np{13.97} & \np{18.0} & \np{111} \\
\# Transitive final & \np{0} & \np{6.5} & \np{25.0} & \np{66.56} & \np{82.5} & \np{515} \\
\end{tabular}
\vspace{-0.25cm}
\end{table}




%% file: figs/chart_bloat_update.tex




\begin{tikzpicture} 
\begin{axis}[xbar,
ymin=0.5,
ymax=2.5,
xmin=0,
enlarge x limits={upper, value=0.2},
width=0.48\textwidth,
height=2.4cm,
bar shift=0pt,
bar width=2.8mm,
axis x line* = bottom,
axis y line* = left,
xtick style={draw=none}, 
ytick style={draw=none}, 
yticklabels = \empty,
xmajorgrids = true,
ticklabel style={
/pgf/number format/fixed,
/pgf/number format/precision=5
}, 
scaled ticks=false,
xlabel=\# Direct dependency updates,
nodes near coords={\pgfmathprintnumber\myy~{\scriptsize(\pgfmathprintnumber\pgfplotspointmeta\%)}},
every node near coord/.append style={
    color = black,
    rotate=0,
    anchor=west
},
visualization depends on=rawx \as \myy,
point meta={x*100/\nbDepDevUpdatesNum},
legend style={
    row sep=3pt,
    draw=none,
    legend columns=2,
    at={(0.5,1.7)},
    anchor=north,
    cells={anchor=west,font=\scriptsize}
},
legend image code/.code={
    \draw[#1, draw=none] (0cm,-0.1cm) rectangle (0.2cm,0.1cm);
},
]

\addplot+[gruen_4a] coordinates {(\nbDepUsedUpdateNum,2)};
\addplot+[rot_9a] coordinates {(\nbDepBloatUpdateNum,1)};

\addlegendentry{Developers updates on used dependencies}
\addlegendentry{Developers updates on bloated dependencies}

\end{axis}
\end{tikzpicture}

%% file: figs/chart_bloat_dependabot_update.tex




\begin{tikzpicture} 
\begin{axis}[xbar,
ymin=0.5,
ymax=2.5,
xmin=0,
enlarge x limits={upper, value=0.2},
width=0.48\textwidth,
height=2.4cm,
bar shift=0pt,
bar width=2.8mm,
axis x line* = bottom,
axis y line* = left,
xtick style={draw=none}, 
ytick style={draw=none}, 
yticklabels = \empty,
xmajorgrids = true,
ticklabel style={
/pgf/number format/fixed,
/pgf/number format/precision=5
}, 
scaled ticks=false,
xlabel=\# Direct dependency updates,
nodes near coords={\pgfmathprintnumber\myy~{\scriptsize(\pgfmathprintnumber\pgfplotspointmeta\%)}},
every node near coord/.append style={
    color = black,
    rotate=0,
    anchor=west
},
visualization depends on=rawx \as \myy,
point meta={x*100/\nbDependabotUpdatesNum},
legend style={
    row sep=3pt,
    draw=none,
    legend columns=2,
    at={(0.5,1.7)},
    anchor=north,
    cells={anchor=west,font=\scriptsize}
},
legend image code/.code={
    \draw[#1, draw=none] (0cm,-0.1cm) rectangle (0.2cm,0.1cm);
},
]

\addplot+[gruen_4a] coordinates {(\nbUsedDependabotUpdateNum,2)};
\addplot+[rot_9a] coordinates {(\nbBloatDependabotUpdateNum,1)};

\addlegendentry{Dependabot updates on used dependencies}
\addlegendentry{Dependabot updates on bloated dependencies}

\end{axis}
\end{tikzpicture}

%% file: main.bbl

\begin{thebibliography}{35}


\ifx \showCODEN    \undefined \def \showCODEN     #1{\unskip}     \fi
\ifx \showDOI      \undefined \def \showDOI       #1{#1}\fi
\ifx \showISBNx    \undefined \def \showISBNx     #1{\unskip}     \fi
\ifx \showISBNxiii \undefined \def \showISBNxiii  #1{\unskip}     \fi
\ifx \showISSN     \undefined \def \showISSN      #1{\unskip}     \fi
\ifx \showLCCN     \undefined \def \showLCCN      #1{\unskip}     \fi
\ifx \shownote     \undefined \def \shownote      #1{#1}          \fi
\ifx \showarticletitle \undefined \def \showarticletitle #1{#1}   \fi
\ifx \showURL      \undefined \def \showURL       {\relax}        \fi
\providecommand\bibfield[2]{#2}
\providecommand\bibinfo[2]{#2}
\providecommand\natexlab[1]{#1}
\providecommand\showeprint[2][]{arXiv:#2}

\bibitem[\protect\citeauthoryear{Agadakos, Demarinis, Jin, Williams-King,
  Alfajardo, Shteinfeld, Williams-King, Kemerlis, and Portokalidis}{Agadakos
  et~al\mbox{.}}{2020}]%
        {Agadakos2020}
\bibfield{author}{\bibinfo{person}{Ioannis Agadakos}, \bibinfo{person}{Nicholas
  Demarinis}, \bibinfo{person}{Di Jin}, \bibinfo{person}{Kent Williams-King},
  \bibinfo{person}{Jearson Alfajardo}, \bibinfo{person}{Benjamin Shteinfeld},
  \bibinfo{person}{David Williams-King}, \bibinfo{person}{Vasileios~P.
  Kemerlis}, {and} \bibinfo{person}{Georgios Portokalidis}.}
  \bibinfo{year}{2020}\natexlab{}.
\newblock \showarticletitle{{Large-scale Debloating of Binary Shared
  Libraries}}.
\newblock \bibinfo{journal}{\emph{Digital Threats: Research and Practice}}
  \bibinfo{volume}{1}, \bibinfo{number}{4} (\bibinfo{year}{2020}),
  \bibinfo{pages}{1--28}.
\newblock
\showISSN{2692-1626}
\urldef\tempurl%
\url{https://doi.org/10.1145/3414997}
\showDOI{\tempurl}


\bibitem[\protect\citeauthoryear{Antignac, Sands, and Schneider}{Antignac
  et~al\mbox{.}}{2017}]%
        {AntignacSS17}
\bibfield{author}{\bibinfo{person}{Thibaud Antignac}, \bibinfo{person}{David
  Sands}, {and} \bibinfo{person}{Gerardo Schneider}.}
  \bibinfo{year}{2017}\natexlab{}.
\newblock \showarticletitle{Data Minimisation: {A} Language-Based Approach}. In
  \bibinfo{booktitle}{\emph{{ICT} Systems Security and Privacy Protection -
  32nd {IFIP} {TC} 11 International Conference, {SEC} 2017, Rome, Italy, May
  29-31, 2017, Proceedings}} \emph{(\bibinfo{series}{{IFIP} Advances in
  Information and Communication Technology})},
  \bibfield{editor}{\bibinfo{person}{Sabrina De~Capitani di~Vimercati} {and}
  \bibinfo{person}{Fabio Martinelli}} (Eds.), Vol.~\bibinfo{volume}{502}.
  \bibinfo{publisher}{Springer}, \bibinfo{pages}{442--456}.
\newblock
\urldef\tempurl%
\url{https://doi.org/10.1007/978-3-319-58469-0\_30}
\showDOI{\tempurl}


\bibitem[\protect\citeauthoryear{Azad, Laperdrix, and Nikiforakis}{Azad
  et~al\mbox{.}}{2019}]%
        {Azad2019}
\bibfield{author}{\bibinfo{person}{Babak~Amin Azad}, \bibinfo{person}{Pierre
  Laperdrix}, {and} \bibinfo{person}{Nick Nikiforakis}.}
  \bibinfo{year}{2019}\natexlab{}.
\newblock \showarticletitle{Less is More: Quantifying the Security Benefits of
  Debloating Web Applications}. In \bibinfo{booktitle}{\emph{Proceedings of the
  28th USENIX Conference on Security Symposium}}
  \emph{(\bibinfo{series}{SEC'19})}. \bibinfo{publisher}{USENIX Association},
  \bibinfo{address}{USA}, \bibinfo{pages}{1697–1714}.
\newblock
\showISBNx{9781939133069}


\bibitem[\protect\citeauthoryear{Boldi and Gousios}{Boldi and Gousios}{2021}]%
        {BoldiG21}
\bibfield{author}{\bibinfo{person}{Paolo Boldi} {and} \bibinfo{person}{Georgios
  Gousios}.} \bibinfo{year}{2021}\natexlab{}.
\newblock \showarticletitle{Fine-Grained Network Analysis for Modern Software
  Ecosystems}.
\newblock \bibinfo{journal}{\emph{{ACM} Trans. Internet Techn.}}
  \bibinfo{volume}{21}, \bibinfo{number}{1} (\bibinfo{year}{2021}),
  \bibinfo{pages}{1:1--1:14}.
\newblock
\urldef\tempurl%
\url{https://doi.org/10.1145/3418209}
\showDOI{\tempurl}


\bibitem[\protect\citeauthoryear{Bruce, Zhang, Arora, Xu, and Kim}{Bruce
  et~al\mbox{.}}{2020}]%
        {Bruce2020}
\bibfield{author}{\bibinfo{person}{Bobby~R. Bruce}, \bibinfo{person}{Tianyi
  Zhang}, \bibinfo{person}{Jaspreet Arora}, \bibinfo{person}{Guoqing~Harry Xu},
  {and} \bibinfo{person}{Miryung Kim}.} \bibinfo{year}{2020}\natexlab{}.
\newblock \showarticletitle{{JShrink: In-depth investigation into debloating
  modern Java applications}}. In \bibinfo{booktitle}{\emph{ESEC/FSE 2020 -
  Proceedings of the 28th ACM Joint Meeting European Software Engineering
  Conference and Symposium on the Foundations of Software Engineering}}.
  \bibinfo{publisher}{Association for Computing Machinery},
  \bibinfo{address}{New York, NY, USA}, \bibinfo{pages}{135--146}.
\newblock
\showISBNx{9781450370431}
\urldef\tempurl%
\url{https://doi.org/10.1145/3368089.3409738}
\showDOI{\tempurl}


\bibitem[\protect\citeauthoryear{Cox}{Cox}{2019}]%
        {Cox2019}
\bibfield{author}{\bibinfo{person}{Russ Cox}.} \bibinfo{year}{2019}\natexlab{}.
\newblock \showarticletitle{{Surviving software dependencies}}.
\newblock \bibinfo{journal}{\emph{Commun. ACM}} \bibinfo{volume}{62},
  \bibinfo{number}{9} (\bibinfo{year}{2019}), \bibinfo{pages}{36--43}.
\newblock
\showISSN{15577317}
\urldef\tempurl%
\url{https://doi.org/10.1145/3347446}
\showDOI{\tempurl}


\bibitem[\protect\citeauthoryear{Dey, Mousavi, Ponce, Fry, Vasilescu,
  Filippova, and Mockus}{Dey et~al\mbox{.}}{2020}]%
        {Dey2020}
\bibfield{author}{\bibinfo{person}{Tapajit Dey}, \bibinfo{person}{Sara
  Mousavi}, \bibinfo{person}{Eduardo Ponce}, \bibinfo{person}{Tanner Fry},
  \bibinfo{person}{Bogdan Vasilescu}, \bibinfo{person}{Anna Filippova}, {and}
  \bibinfo{person}{Audris Mockus}.} \bibinfo{year}{2020}\natexlab{}.
\newblock \showarticletitle{{Detecting and Characterizing Bots that Commit
  Code}}. In \bibinfo{booktitle}{\emph{Proceedings - 2020 IEEE/ACM 17th
  International Conference on Mining Software Repositories, MSR 2020}}.
  \bibinfo{publisher}{Association for Computing Machinery},
  \bibinfo{address}{New York, NY, USA}, \bibinfo{pages}{209--219}.
\newblock
\showISBNx{9781450379571}
\urldef\tempurl%
\url{https://doi.org/10.1145/3379597.3387478}
\showDOI{\tempurl}
\showeprint[arxiv]{2003.03172}


\bibitem[\protect\citeauthoryear{Durieux, Soto{-}Valero, and Baudry}{Durieux
  et~al\mbox{.}}{2021}]%
        {durieux21}
\bibfield{author}{\bibinfo{person}{Thomas Durieux},
  \bibinfo{person}{C{\'{e}}sar Soto{-}Valero}, {and} \bibinfo{person}{Benoit
  Baudry}.} \bibinfo{year}{2021}\natexlab{}.
\newblock \showarticletitle{{DUETS:} {A} Dataset of Reproducible Pairs of Java
  Library-Clients}. In \bibinfo{booktitle}{\emph{IEEE International Working
  Conference on Mining Software Repositories}}.
\newblock


\bibitem[\protect\citeauthoryear{Erlenhov, Neto, and Leitner}{Erlenhov
  et~al\mbox{.}}{2020}]%
        {Erlenhov20}
\bibfield{author}{\bibinfo{person}{Linda Erlenhov}, \bibinfo{person}{Francisco
  Gomes De~Oliveira Neto}, {and} \bibinfo{person}{Philipp Leitner}.}
  \bibinfo{year}{2020}\natexlab{}.
\newblock \bibinfo{booktitle}{\emph{{An empirical study of bots in software
  development: Characteristics and challenges from a practitioner's
  perspective}}}.
\newblock \bibinfo{publisher}{Association for Computing Machinery},
  \bibinfo{address}{New York, NY, USA}, \bibinfo{pages}{445--455}.
\newblock
\showISBNx{9781450370431}
\urldef\tempurl%
\url{https://doi.org/10.1145/3368089.3409680}
\showDOI{\tempurl}
\showeprint[arxiv]{2005.13969}


\bibitem[\protect\citeauthoryear{Gustavsson}{Gustavsson}{2020}]%
        {Gustavsson20}
\bibfield{author}{\bibinfo{person}{Tomas Gustavsson}.}
  \bibinfo{year}{2020}\natexlab{}.
\newblock \showarticletitle{Managing the Open Source Dependency}.
\newblock \bibinfo{journal}{\emph{Computer}} \bibinfo{volume}{53},
  \bibinfo{number}{2} (\bibinfo{year}{2020}), \bibinfo{pages}{83--87}.
\newblock
\urldef\tempurl%
\url{https://doi.org/10.1109/MC.2019.2955869}
\showDOI{\tempurl}


\bibitem[\protect\citeauthoryear{Hague, Lin, and Hong}{Hague
  et~al\mbox{.}}{2019}]%
        {HagueLH19}
\bibfield{author}{\bibinfo{person}{Matthew Hague}, \bibinfo{person}{Anthony~W.
  Lin}, {and} \bibinfo{person}{Chih{-}Duo Hong}.}
  \bibinfo{year}{2019}\natexlab{}.
\newblock \showarticletitle{{CSS} Minification via Constraint Solving}.
\newblock \bibinfo{journal}{\emph{{ACM} Trans. Program. Lang. Syst.}}
  \bibinfo{volume}{41}, \bibinfo{number}{2} (\bibinfo{year}{2019}),
  \bibinfo{pages}{12:1--12:76}.
\newblock
\urldef\tempurl%
\url{https://doi.org/10.1145/3310337}
\showDOI{\tempurl}


\bibitem[\protect\citeauthoryear{Hilton, Bell, and Marinov}{Hilton
  et~al\mbox{.}}{2018}]%
        {Hilton0M18}
\bibfield{author}{\bibinfo{person}{Michael Hilton}, \bibinfo{person}{Jonathan
  Bell}, {and} \bibinfo{person}{Darko Marinov}.}
  \bibinfo{year}{2018}\natexlab{}.
\newblock \showarticletitle{A large-scale study of test coverage evolution}. In
  \bibinfo{booktitle}{\emph{Proceedings of ASE}}. \bibinfo{publisher}{{ACM}},
  \bibinfo{pages}{53--63}.
\newblock


\bibitem[\protect\citeauthoryear{Holzmann}{Holzmann}{2015}]%
        {Holzmann2015}
\bibfield{author}{\bibinfo{person}{Gerard~J. Holzmann}.}
  \bibinfo{year}{2015}\natexlab{}.
\newblock \showarticletitle{{Code inflation}}.
\newblock \bibinfo{journal}{\emph{IEEE Software}} \bibinfo{volume}{32},
  \bibinfo{number}{2} (\bibinfo{date}{March} \bibinfo{year}{2015}),
  \bibinfo{pages}{10--13}.
\newblock
\showISSN{07407459}
\urldef\tempurl%
\url{https://doi.org/10.1109/MS.2015.40}
\showDOI{\tempurl}


\bibitem[\protect\citeauthoryear{Jiang, Wu, and Liu}{Jiang
  et~al\mbox{.}}{2016}]%
        {Jiang2016}
\bibfield{author}{\bibinfo{person}{Yufei Jiang}, \bibinfo{person}{Dinghao Wu},
  {and} \bibinfo{person}{Peng Liu}.} \bibinfo{year}{2016}\natexlab{}.
\newblock \showarticletitle{{JRed: Program Customization and Bloatware
  Mitigation Based on Static Analysis}}. In
  \bibinfo{booktitle}{\emph{Proceedings - International Computer Software and
  Applications Conference}}, Vol.~\bibinfo{volume}{1}.
  \bibinfo{publisher}{{IEEE} Press.}, \bibinfo{address}{New York},
  \bibinfo{pages}{12--21}.
\newblock
\showISBNx{9781467388450}
\showISSN{07303157}
\urldef\tempurl%
\url{https://doi.org/10.1109/COMPSAC.2016.146}
\showDOI{\tempurl}


\bibitem[\protect\citeauthoryear{Koo, Ghavamnia, and Polychronakis}{Koo
  et~al\mbox{.}}{2019}]%
        {Koo2019}
\bibfield{author}{\bibinfo{person}{Hyungjoon Koo}, \bibinfo{person}{Seyedhamed
  Ghavamnia}, {and} \bibinfo{person}{Michalis Polychronakis}.}
  \bibinfo{year}{2019}\natexlab{}.
\newblock \showarticletitle{Configuration-Driven Software Debloating}. In
  \bibinfo{booktitle}{\emph{Proceedings of the 12th European Workshop on
  Systems Security}} \emph{(\bibinfo{series}{EuroSec '19})}.
  \bibinfo{publisher}{Association for Computing Machinery},
  \bibinfo{address}{New York, NY, USA}, Article \bibinfo{articleno}{9},
  \bibinfo{numpages}{6}~pages.
\newblock
\showISBNx{9781450362740}
\urldef\tempurl%
\url{https://doi.org/10.1145/3301417.3312501}
\showDOI{\tempurl}


\bibitem[\protect\citeauthoryear{Kula, German, Ouni, Ishio, and Inoue}{Kula
  et~al\mbox{.}}{2018}]%
        {Kula2018}
\bibfield{author}{\bibinfo{person}{Raula~Gaikovina Kula},
  \bibinfo{person}{Daniel~M. German}, \bibinfo{person}{Ali Ouni},
  \bibinfo{person}{Takashi Ishio}, {and} \bibinfo{person}{Katsuro Inoue}.}
  \bibinfo{year}{2018}\natexlab{}.
\newblock \showarticletitle{{Do developers update their library dependencies?:
  An empirical study on the impact of security advisories on library
  migration}}.
\newblock \bibinfo{journal}{\emph{Empirical Software Engineering}}
  \bibinfo{volume}{23}, \bibinfo{number}{1} (\bibinfo{date}{Feb.}
  \bibinfo{year}{2018}), \bibinfo{pages}{384--417}.
\newblock
\showISSN{15737616}
\urldef\tempurl%
\url{https://doi.org/10.1007/s10664-017-9521-5}
\showDOI{\tempurl}
\showeprint[arxiv]{1709.04621}


\bibitem[\protect\citeauthoryear{Landman, Serebrenik, and Vinju}{Landman
  et~al\mbox{.}}{2017}]%
        {Landman2017}
\bibfield{author}{\bibinfo{person}{Davy Landman}, \bibinfo{person}{Alexander
  Serebrenik}, {and} \bibinfo{person}{Jurgen~J. Vinju}.}
  \bibinfo{year}{2017}\natexlab{}.
\newblock \showarticletitle{Challenges for Static Analysis of Java Reflection:
  Literature Review and Empirical Study}. In
  \bibinfo{booktitle}{\emph{Proceedings of the 39th International Conference on
  Software Engineering}} \emph{(\bibinfo{series}{ICSE '17})}.
  \bibinfo{publisher}{{IEEE} Press.}, \bibinfo{address}{New York},
  \bibinfo{pages}{507–518}.
\newblock
\showISBNx{9781538638682}
\urldef\tempurl%
\url{https://doi.org/10.1109/ICSE.2017.53}
\showDOI{\tempurl}


\bibitem[\protect\citeauthoryear{Livshits, Sridharan, Smaragdakis, Lhot\'{a}k,
  Amaral, Chang, Guyer, Khedker, M\o{}ller, and Vardoulakis}{Livshits
  et~al\mbox{.}}{2015}]%
        {Livshits2015}
\bibfield{author}{\bibinfo{person}{Benjamin Livshits}, \bibinfo{person}{Manu
  Sridharan}, \bibinfo{person}{Yannis Smaragdakis}, \bibinfo{person}{Ond\v{r}ej
  Lhot\'{a}k}, \bibinfo{person}{J.~Nelson Amaral},
  \bibinfo{person}{Bor-Yuh~Evan Chang}, \bibinfo{person}{Samuel~Z. Guyer},
  \bibinfo{person}{Uday~P. Khedker}, \bibinfo{person}{Anders M\o{}ller}, {and}
  \bibinfo{person}{Dimitrios Vardoulakis}.} \bibinfo{year}{2015}\natexlab{}.
\newblock \showarticletitle{In Defense of Soundiness: A Manifesto}.
\newblock \bibinfo{journal}{\emph{Commun. ACM}} \bibinfo{volume}{58},
  \bibinfo{number}{2} (\bibinfo{date}{Jan.} \bibinfo{year}{2015}),
  \bibinfo{pages}{44–46}.
\newblock
\showISSN{0001-0782}
\urldef\tempurl%
\url{https://doi.org/10.1145/2644805}
\showDOI{\tempurl}


\bibitem[\protect\citeauthoryear{Loriot, Madeiral, and Monperrus}{Loriot
  et~al\mbox{.}}{2020}]%
        {styler}
\bibfield{author}{\bibinfo{person}{Benjamin Loriot}, \bibinfo{person}{Fernanda
  Madeiral}, {and} \bibinfo{person}{Martin Monperrus}.}
  \bibinfo{year}{2020}\natexlab{}.
\newblock \showarticletitle{{Styler: Learning formatting conventions to repair
  checkstyle errors}}.
\newblock \bibinfo{journal}{\emph{arXiv}}  \bibinfo{volume}{1}
  (\bibinfo{year}{2020}), \bibinfo{pages}{1--1}.
\newblock
\showISSN{23318422}
\showeprint[arxiv]{1904.01754}


\bibitem[\protect\citeauthoryear{Macias, Mathur, Bruce, Zhang, and Kim}{Macias
  et~al\mbox{.}}{2020}]%
        {Macias2020}
\bibfield{author}{\bibinfo{person}{Konner Macias}, \bibinfo{person}{Mihir
  Mathur}, \bibinfo{person}{Bobby~R. Bruce}, \bibinfo{person}{Tianyi Zhang},
  {and} \bibinfo{person}{Miryung Kim}.} \bibinfo{year}{2020}\natexlab{}.
\newblock \showarticletitle{WebJShrink: A Web Service for Debloating Java
  Bytecode}. In \bibinfo{booktitle}{\emph{Proceedings of ESEC/FSE}}.
  \bibinfo{pages}{1665–1669}.
\newblock


\bibitem[\protect\citeauthoryear{Qian, Hu, Alharthi, Chung, Kim, and Lee}{Qian
  et~al\mbox{.}}{2019}]%
        {Qian2019}
\bibfield{author}{\bibinfo{person}{Chenxiong Qian}, \bibinfo{person}{Hong Hu},
  \bibinfo{person}{Mansour Alharthi}, \bibinfo{person}{Pak~Ho Chung},
  \bibinfo{person}{Taesoo Kim}, {and} \bibinfo{person}{Wenke Lee}.}
  \bibinfo{year}{2019}\natexlab{}.
\newblock \showarticletitle{RAZOR: A Framework for Post-Deployment Software
  Debloating}. In \bibinfo{booktitle}{\emph{Proceedings of the 28th USENIX
  Conference on Security Symposium}} \emph{(\bibinfo{series}{SEC'19})}.
  \bibinfo{publisher}{USENIX Association}, \bibinfo{address}{USA},
  \bibinfo{pages}{1733–1750}.
\newblock
\showISBNx{9781939133069}


\bibitem[\protect\citeauthoryear{Qian, Koo, Oh, Kim, and Lee}{Qian
  et~al\mbox{.}}{2020}]%
        {Qian2020}
\bibfield{author}{\bibinfo{person}{Chenxiong Qian}, \bibinfo{person}{Hyungjoon
  Koo}, \bibinfo{person}{Chang~Seok Oh}, \bibinfo{person}{Taesoo Kim}, {and}
  \bibinfo{person}{Wenke Lee}.} \bibinfo{year}{2020}\natexlab{}.
\newblock \showarticletitle{{Slimium: Debloating the Chromium Browser with
  Feature Subsetting}}. In \bibinfo{booktitle}{\emph{Proceedings of the ACM
  Conference on Computer and Communications Security}}.
  \bibinfo{publisher}{Association for Computing Machinery},
  \bibinfo{address}{New York, NY, USA}, \bibinfo{pages}{461--476}.
\newblock
\showISBNx{9781450370899}
\showISSN{15437221}
\urldef\tempurl%
\url{https://doi.org/10.1145/3372297.3417866}
\showDOI{\tempurl}


\bibitem[\protect\citeauthoryear{Quach, Prakash, and Yan}{Quach
  et~al\mbox{.}}{2018}]%
        {Quach2018}
\bibfield{author}{\bibinfo{person}{Anh Quach}, \bibinfo{person}{Aravind
  Prakash}, {and} \bibinfo{person}{Lok Yan}.} \bibinfo{year}{2018}\natexlab{}.
\newblock \showarticletitle{Debloating Software through Piece-Wise Compilation
  and Loading}. In \bibinfo{booktitle}{\emph{Proceedings of the 27th USENIX
  Conference on Security Symposium}} \emph{(\bibinfo{series}{SEC'18})}.
  \bibinfo{publisher}{USENIX Association}, \bibinfo{address}{USA},
  \bibinfo{pages}{869–886}.
\newblock
\showISBNx{9781931971461}
\urldef\tempurl%
\url{https://doi.org/10.5555/3277203.3277269}
\showDOI{\tempurl}


\bibitem[\protect\citeauthoryear{Rastogi, Davidson, {De Carli}, Jha, and
  McDaniel}{Rastogi et~al\mbox{.}}{2017}]%
        {Rastogi2017}
\bibfield{author}{\bibinfo{person}{Vaibhav Rastogi}, \bibinfo{person}{Drew
  Davidson}, \bibinfo{person}{Lorenzo {De Carli}}, \bibinfo{person}{Somesh
  Jha}, {and} \bibinfo{person}{Patrick McDaniel}.}
  \bibinfo{year}{2017}\natexlab{}.
\newblock \showarticletitle{{Cimplifier: Automatically debloating containers}}.
  In \bibinfo{booktitle}{\emph{Proceedings of the ACM SIGSOFT Symposium on the
  Foundations of Software Engineering}}, Vol.~\bibinfo{volume}{Part F130154}.
  \bibinfo{publisher}{Association for Computing Machinery},
  \bibinfo{address}{New York, NY, USA}, \bibinfo{pages}{476--486}.
\newblock
\showISBNx{9781450351058}
\urldef\tempurl%
\url{https://doi.org/10.1145/3106237.3106271}
\showDOI{\tempurl}


\bibitem[\protect\citeauthoryear{Reif, K\"{u}bler, Eichberg, and Mezini}{Reif
  et~al\mbox{.}}{2018}]%
        {Reif2018}
\bibfield{author}{\bibinfo{person}{Michael Reif}, \bibinfo{person}{Florian
  K\"{u}bler}, \bibinfo{person}{Michael Eichberg}, {and} \bibinfo{person}{Mira
  Mezini}.} \bibinfo{year}{2018}\natexlab{}.
\newblock \showarticletitle{Systematic Evaluation of the Unsoundness of Call
  Graph Construction Algorithms for Java}. In
  \bibinfo{booktitle}{\emph{Companion Proceedings for the ISSTA/ECOOP 2018
  Workshops}} \emph{(\bibinfo{series}{ISSTA '18})}.
  \bibinfo{publisher}{Association for Computing Machinery},
  \bibinfo{address}{New York, NY, USA}, \bibinfo{pages}{107–112}.
\newblock
\showISBNx{9781450359399}
\urldef\tempurl%
\url{https://doi.org/10.1145/3236454.3236503}
\showDOI{\tempurl}


\bibitem[\protect\citeauthoryear{Sharif, Gehani, Abubakar, and Zaffar}{Sharif
  et~al\mbox{.}}{2018}]%
        {Sharif2018}
\bibfield{author}{\bibinfo{person}{Hashim Sharif}, \bibinfo{person}{Ashish
  Gehani}, \bibinfo{person}{Muhammad Abubakar}, {and} \bibinfo{person}{Fareed
  Zaffar}.} \bibinfo{year}{2018}\natexlab{}.
\newblock \showarticletitle{{Trimmer: Application specialization for code
  debloating}}. In \bibinfo{booktitle}{\emph{ASE 2018 - Proceedings of the 33rd
  ACM/IEEE International Conference on Automated Software Engineering}}
  \emph{(\bibinfo{series}{ASE 2018})}. \bibinfo{publisher}{Association for
  Computing Machinery}, \bibinfo{address}{New York, NY, USA},
  \bibinfo{pages}{329--339}.
\newblock
\showISBNx{9781450359375}
\urldef\tempurl%
\url{https://doi.org/10.1145/3238147.3238160}
\showDOI{\tempurl}


\bibitem[\protect\citeauthoryear{Soto-Valero, Benelallam, Harrand, Barais, and
  Baudry}{Soto-Valero et~al\mbox{.}}{2019}]%
        {Valero2019}
\bibfield{author}{\bibinfo{person}{Cesar Soto-Valero}, \bibinfo{person}{Amine
  Benelallam}, \bibinfo{person}{Nicolas Harrand}, \bibinfo{person}{Olivier
  Barais}, {and} \bibinfo{person}{Benoit Baudry}.}
  \bibinfo{year}{2019}\natexlab{}.
\newblock \showarticletitle{{The emergence of software diversity in maven
  central}}. In \bibinfo{booktitle}{\emph{IEEE International Working Conference
  on Mining Software Repositories}} \emph{(\bibinfo{series}{MSR '19})},
  Vol.~\bibinfo{volume}{2019-May}. \bibinfo{publisher}{{IEEE} Press.},
  \bibinfo{address}{New York}, \bibinfo{pages}{333--343}.
\newblock
\showISBNx{9781728134123}
\showISSN{21601860}
\urldef\tempurl%
\url{https://doi.org/10.1109/MSR.2019.00059}
\showDOI{\tempurl}
\showeprint[arxiv]{1903.05394}


\bibitem[\protect\citeauthoryear{{Soto-Valero}, {Durieux}, {Harrand}, and
  {Baudry}}{{Soto-Valero} et~al\mbox{.}}{2020}]%
        {Valero2020traces}
\bibfield{author}{\bibinfo{person}{C{\'e}sar {Soto-Valero}},
  \bibinfo{person}{Thomas {Durieux}}, \bibinfo{person}{Nicolas {Harrand}},
  {and} \bibinfo{person}{Benoit {Baudry}}.} \bibinfo{year}{2020}\natexlab{}.
\newblock \showarticletitle{{Trace-based Debloat for Java Bytecode}}.
\newblock \bibinfo{journal}{\emph{arXiv}}  \bibinfo{volume}{1}, Article
  \bibinfo{articleno}{arXiv:2008.08401} (\bibinfo{date}{Aug.}
  \bibinfo{year}{2020}), \bibinfo{numpages}{12}~pages.
\newblock
\showeprint[arxiv]{cs.SE/2008.08401}


\bibitem[\protect\citeauthoryear{Soto-Valero, Harrand, Monperrus, and
  Baudry}{Soto-Valero et~al\mbox{.}}{2021}]%
        {Valero2020}
\bibfield{author}{\bibinfo{person}{C{\'{e}}sar Soto-Valero},
  \bibinfo{person}{Nicolas Harrand}, \bibinfo{person}{Martin Monperrus}, {and}
  \bibinfo{person}{Benoit Baudry}.} \bibinfo{year}{2021}\natexlab{}.
\newblock \showarticletitle{{A Comprehensive Study of Bloated Dependencies in
  the Maven Ecosystem}}.
\newblock \bibinfo{journal}{\emph{Empirical Software Engineering}}
  \bibinfo{volume}{26}, \bibinfo{number}{3} (\bibinfo{year}{2021}),
  \bibinfo{pages}{1--44}.
\newblock


\bibitem[\protect\citeauthoryear{Spinellis}{Spinellis}{2017}]%
        {Spinellis17}
\bibfield{author}{\bibinfo{person}{Diomidis Spinellis}.}
  \bibinfo{year}{2017}\natexlab{}.
\newblock \showarticletitle{A repository of Unix history and evolution}.
\newblock \bibinfo{journal}{\emph{Empir. Softw. Eng.}} \bibinfo{volume}{22},
  \bibinfo{number}{3} (\bibinfo{year}{2017}), \bibinfo{pages}{1372--1404}.
\newblock
\urldef\tempurl%
\url{https://doi.org/10.1007/s10664-016-9445-5}
\showDOI{\tempurl}


\bibitem[\protect\citeauthoryear{Sun, Li, Zhang, Gu, and Su}{Sun
  et~al\mbox{.}}{2018}]%
        {Sun2018}
\bibfield{author}{\bibinfo{person}{Chengnian Sun}, \bibinfo{person}{Yuanbo Li},
  \bibinfo{person}{Qirun Zhang}, \bibinfo{person}{Tianxiao Gu}, {and}
  \bibinfo{person}{Zhendong Su}.} \bibinfo{year}{2018}\natexlab{}.
\newblock \showarticletitle{Perses: Syntax-Guided Program Reduction}. In
  \bibinfo{booktitle}{\emph{Proceedings of the 40th International Conference on
  Software Engineering}} \emph{(\bibinfo{series}{ICSE '18})}.
  \bibinfo{publisher}{Association for Computing Machinery},
  \bibinfo{address}{New York, NY, USA}, \bibinfo{pages}{361–371}.
\newblock
\showISBNx{9781450356381}
\urldef\tempurl%
\url{https://doi.org/10.1145/3180155.3180236}
\showDOI{\tempurl}


\bibitem[\protect\citeauthoryear{Teyton, Falleri, Palyart, and Blanc}{Teyton
  et~al\mbox{.}}{2014}]%
        {TeytonFPB14}
\bibfield{author}{\bibinfo{person}{C{\'{e}}dric Teyton},
  \bibinfo{person}{Jean{-}R{\'{e}}my Falleri}, \bibinfo{person}{Marc Palyart},
  {and} \bibinfo{person}{Xavier Blanc}.} \bibinfo{year}{2014}\natexlab{}.
\newblock \showarticletitle{A study of library migrations in Java}.
\newblock \bibinfo{journal}{\emph{J. Softw. Evol. Process.}}
  \bibinfo{volume}{26}, \bibinfo{number}{11} (\bibinfo{year}{2014}),
  \bibinfo{pages}{1030--1052}.
\newblock
\urldef\tempurl%
\url{https://doi.org/10.1002/smr.1660}
\showDOI{\tempurl}


\bibitem[\protect\citeauthoryear{V{\'{a}}zquez, Bergel, Vidal, {D{\'{i}}az
  Pace}, and Marcos}{V{\'{a}}zquez et~al\mbox{.}}{2019}]%
        {Vazquez2019}
\bibfield{author}{\bibinfo{person}{H.~C. V{\'{a}}zquez}, \bibinfo{person}{A.
  Bergel}, \bibinfo{person}{S. Vidal}, \bibinfo{person}{J.~A. {D{\'{i}}az
  Pace}}, {and} \bibinfo{person}{C. Marcos}.} \bibinfo{year}{2019}\natexlab{}.
\newblock \showarticletitle{{Slimming javascript applications: An approach for
  removing unused functions from javascript libraries}}.
\newblock \bibinfo{journal}{\emph{Information and Software Technology}}
  \bibinfo{volume}{107} (\bibinfo{year}{2019}), \bibinfo{pages}{18--29}.
\newblock
\showISSN{09505849}
\urldef\tempurl%
\url{https://doi.org/10.1016/j.infsof.2018.10.009}
\showDOI{\tempurl}


\bibitem[\protect\citeauthoryear{Wessel and Steinmacher}{Wessel and
  Steinmacher}{2020}]%
        {wessel2020inconvenient}
\bibfield{author}{\bibinfo{person}{Mairieli Wessel} {and} \bibinfo{person}{Igor
  Steinmacher}.} \bibinfo{year}{2020}\natexlab{}.
\newblock \showarticletitle{The Inconvenient Side of Software Bots on Pull
  Requests}. In \bibinfo{booktitle}{\emph{Proceedings of the IEEE/ACM 42nd
  International Conference on Software Engineering Workshops}}
  \emph{(\bibinfo{series}{ICSEW'20})}. \bibinfo{publisher}{Association for
  Computing Machinery}, \bibinfo{address}{New York, NY, USA},
  \bibinfo{pages}{51–55}.
\newblock
\showISBNx{9781450379632}
\urldef\tempurl%
\url{https://doi.org/10.1145/3387940.3391504}
\showDOI{\tempurl}


\bibitem[\protect\citeauthoryear{Xin, Kim, Zhang, and Orso}{Xin
  et~al\mbox{.}}{2020}]%
        {Xin2020}
\bibfield{author}{\bibinfo{person}{Qi Xin}, \bibinfo{person}{Myeongsoo Kim},
  \bibinfo{person}{Qirun Zhang}, {and} \bibinfo{person}{Alessandro Orso}.}
  \bibinfo{year}{2020}\natexlab{}.
\newblock \showarticletitle{Program Debloating via Stochastic Optimization}. In
  \bibinfo{booktitle}{\emph{Proceedings of the ACM/IEEE 42nd International
  Conference on Software Engineering: New Ideas and Emerging Results}}
  \emph{(\bibinfo{series}{ICSE-NIER '20})}. \bibinfo{publisher}{Association for
  Computing Machinery}, \bibinfo{address}{New York, NY, USA},
  \bibinfo{pages}{65–68}.
\newblock
\showISBNx{9781450371261}
\urldef\tempurl%
\url{https://doi.org/10.1145/3377816.3381739}
\showDOI{\tempurl}


\end{thebibliography}
